\documentclass[a4paper, 11 pt]{article}
\usepackage[T1]{fontenc}
\usepackage[utf8]{inputenc}
\usepackage[english]{babel}
\usepackage{amsmath}
\usepackage{amssymb}
\usepackage{mathtools}
\usepackage{dsfont}
\usepackage{subcaption}
\usepackage{booktabs}
\usepackage{setspace}
\usepackage{cite}
\usepackage{verbatim}

\usepackage{amstext} 
\usepackage{array}   
\usepackage{geometry}
\usepackage[usenames,dvipsnames]{color}
\usepackage{hyperref}
\hypersetup{
  colorlinks,
  citecolor=Blue,
 linkcolor=Blue,
  urlcolor=Blue}

\numberwithin{equation}{section}

\def\be{\begin{equation}}
\def\ee{\end{equation}}
\def\rme{{e}}
\newcommand{\nn}{\nonumber}
\newcommand{\diff}{\mathrm{d}}
\newcommand{\ii}{\mathrm{i}} 

\def\FF{\mathsf{F}}
\def\kRRR{k_{RRR}}
\def\kR{k_R}

\newcommand{\ndt}{\noindent} 
\renewcommand{\=}{\; = \;} 
\newcommand{\CN}{\mathcal{N}}

\begin{document}

\pagestyle{empty}

\begin{center}
$\,$
\vskip 1cm

{\LARGE{\textsf{\textbf{Quantum black holes:\\[3mm]  supersymmetry and exact results}}  }}

\vskip 1cm

{
{\bf Davide Cassani,$^{\,\textrm a}\;$
Sameer Murthy$^{\,\textrm b}$}
}

\end{center}

\vskip 0.5cm

\renewcommand{\thefootnote}{\arabic{footnote}}

\begin{center}
$^{\textrm a}$ {\it INFN, Sezione di Padova, Via Marzolo 8, 35131 Padova, Italy},\\ [2mm] 
$^{\textrm b}${\it Department of Mathematics, King's College London,\\
The Strand, London WC2R 2LS, U.K.}

\end{center}

\vskip 2cm

\begin{abstract}

\noindent      The explanation of black hole entropy as statistical entropy is one of the big successes of string theory. 
    In this article we review recent progress in this subject, focussing on understanding quantum effects on black hole entropy. 
    Supersymmetry plays a key role in these developments and leads to prototype systems where we can discuss quantum effects to great precision.    
    Our discussion has two strands, both of which involve the gravitational path integral that calculates the supersymmetric index.
    In the first strand we discuss supersymmetric black holes in the microcanonical ensemble, which are decoupled from the environment and can be treated as 
    independent quantum systems. Using methods of supersymmetric localization one can arrive at the integer quantum degeneracies of such systems purely in terms of the gravitational variables. 
    In the second strand we consider grand-canonical ensembles in gravity in which black holes arise as a finite-action excitation in asymptotically flat or Anti de Sitter space. 
    In this context we discuss the saddle-points of the gravitational index, and how they reproduce the black hole action and entropy 
    that agree with the index of the holographic superconformal field theory even beyond the leading order in the semiclassical approximation. 
    Finally we discuss how the gravitational index informs us about the detailed structure of the non-perturbative sum over saddle points and the resulting phases of the theory. 
    Throughout, we try to highlight the central role played by the methods as well as foundational concepts of supergravity in driving these developments.

\vskip 2cm
{\it \noindent Invited contribution to the book ``Half a Century of Supergravity'', editors A.~Ceresole and G.~Dall'Agata.}

\end{abstract}

%
\newpage
\setcounter{page}{1}
\pagestyle{plain}

\tableofcontents

\vskip 1cm

\section{Introduction and summary of the article}

Einstein's General Relativity (GR) minimally coupled to matter fields can be regarded as an 
effective theory arising in the low-energy limit from the fundamental theory of quantum gravity. 
In the semiclassical version of this theory, it was discovered in the 1970s that 
Black Holes (BH) display thermodynamical properties~\cite{Bekenstein:1973ur,Hawking:1975vcx}.  
In particular, the thermodynamic entropy of a black hole is given by the Bekenstein-Hawking area law, 
which in four dimensions is 
\be \label{eq:BekHaw}
S_\text{BH} \= k_B \,\frac{A_\text{H}\, c^3}{4 \, G \, \hbar} \,,
\ee
where~$A_\text{H}$ is the area of the BH event horizon 
This appearance in this formula of the fundamental constants: the speed of light~$c$, 
Planck's constant~$\hbar$, 
Newton's gravitational constant~$G$, 
and the Boltzmann constant~$k_B$,
shows a deep connection between gravity,
quantum mechanics, and thermodynamics. 
(We often suppress these constants in the presentation  below.)
For a BH of solar mass, the entropy has the astonishingly large value 
$S_\text{BH}\approx 10^{77} \, k_B$.
The area law~$S=A/4G$ is very general and applies to any BH in any dimension in the thermodynamic, semi-classical limit when the BH is very large. 

The Bekenstein-Hawking formula and its refinements that we discuss below 
provide an important quantitative test for the fundamental quantum theory of gravity. 
In the fundamental theory the thermodynamic entropy should have a statistical origin in 
terms of a large number of microscopic states (\emph{microstates}), 
just like the entropy of the gas in a room arises from the fact that it exists in a
large number of translational, vibrational, and rotational microstates of its molecules. 

\vskip 4pt

One of the biggest successes of string theory is the explanation of the statistical entropy of a BH, 
that appeared in the 1990s in the breakthrough work of Strominger-Vafa~\cite{Strominger:1996sh}.
For a particular type of supersymmetric BH in string theory compactified to 
five-dimensional asymptotically flat space, 
this work explained the Bekenstein-Hawking entropy as the statistical entropy of the 
microstates of weakly-coupled wrapped D-branes with the same quantum numbers in string theory as that of the BH. 
These two pictures of a BH in string theory are related by changing the string coupling  constant. 
Due to supersymmetry, the weak coupling result is protected by quantum 
corrections and can be extrapolated all the way to strong coupling. 
These breakthroughs led to many developments in the subject in different directions, 
see e.g.~\cite{Mohaupt:2000mj,Sen:2007qy,Sen:2008vm} for a review of some aspects of  the story.  

\vskip 4pt

In this article we review more recent progress on \emph{quantum} effects on understanding the black hole entropy. 
To motivate this, we observe that the universality of the thermodynamic Bekenstein-Hawking area law is also 
self-limiting. Indeed, in order to access details of the microscopic theory of quantum gravity, we need to go beyond the semiclassical limit and probe the quantum effects 
that become relevant only when the finite size of the black hole horizon is taken into account.
Our emphasis here is not so much to obtain  realistic features of string compactifications 
as to obtain models of quantum gravity where we can have a detailed explanation of 
fundamental quantum features of the theory. 
As we explain, in the context of supersymmetric BHs in string theory, one has been able to push these models and explanations remarkably far. 

\vskip 4pt

Supergravity provides the natural environment in which to study supersymmetric black holes.
One well-known role of supergravity in the developments above is to provide consistent low-energy descriptions of
larger UV-complete theories of gravity. 
Solutions of supergravity can therefore be embedded in such a UV theory. 
Our attempt in this article is to highlight, in addition, 
a more subtle but important aspect of the theory: some of the deep ideas that underlie  
the construction of the supergravity formalism, like off-shell (conformal) supergravity, the meaning of gauge-invariance, 
the construction of Killing spinors, 
the constrained quantum corrections,
play an important conceptual as well as technical role in the evaluation of the Gravitational Path Integral (GPI) at the  quantum level.

\vskip 2pt

We divide the discussion of quantum effects broadly into two strands, namely the microcanonical and grand-canonical ensembles,
which naturally arise from the initial thermodynamic motivation.
Although this division helps for presentational purposes, these two strands are 
actually unified under the umbrella of the AdS/CFT correspondence, as we see below.

\vskip 6pt

\ndt {\bf Exact quantum entropy and microcanonical ensemble.} 
The first strand, roughly speaking, addresses the question ``What are the molecules of a BH?''.  
Thinking of the BH as a quantum-statistical system, 
we associate a Hilbert space~$\mathcal{H}_\text{BH}$ to it,
and the entropy is given in terms of its dimension through the 
Boltzmann equation~$S_\text{BH} = \log \text{dim}(\mathcal{H}_\text{BH})$.
Combining this with the Bekenstein-Hawking entropy formula 
leads to the \emph{Boltzmann equation for quantum black holes}, 
\be \label{eq:BoltzBH}
\log \text{dim}(\mathcal{H}_\text{BH}) \= \frac{A_\text{H}}{4 \, G} + \dots\,, 
\ee
where both sides of the equation~\eqref{eq:BoltzBH} are functions of the conserved quantum numbers of the BH. 
The ellipsis on the right-hand side of~\eqref{eq:BoltzBH} represents quantities 
that are ignored in the thermodynamic limit when the size of the BH is large. 

If we can calculate and sum up quantum corrections to the entropy, we would obtain  
quantitative access to the simplest signature of the underlying quantum system, namely the dimension 
of its Hilbert space.
This is an important macroscopic window into 
the microscopic theory of quantum gravity.
However, there is an immediate problem with this idea. 
The left-hand side of the formula~\eqref{eq:BoltzBH} is a precise number which is the logarithm of 
an integer that counts the number of possible states of all the degrees of freedom of the quantum system in the 
microcanonical ensemble.
On the other hand, the phenomenon of Hawking radiation means that we cannot isolate or classify  
the quanta surrounding a BH as ``belonging to'' versus ``being outside'' a BH. 

\vskip 2pt

An opportunity for progress is provided by supersymmetric BHs because 
they have zero temperature and therefore do not radiate.\footnote{One may think that all extremal black holes---not only 
supersymmetric ones---do not radiate, 
and that they can have a large zero-temperature Bekenstein-Hawking entropy. 
However, recent developments \cite{Iliesiu:2020qvm,Iliesiu:2022kny} indicate that unless protected by supersymmetry, 
the macroscopic ground state degeneracy suggested by the area of the horizon in the semiclassical theory is generically lifted by strong quantum effects,  
so that the quantum entropy of extremal non-supersymmetric black holes is generically expected to vanish.}  
At least at the semi-classical level, they are isolated systems where 
we can study the Boltzmann equation to great accuracy. 
In particular, Equation~\eqref{eq:BoltzBH}  leads to many sharp questions: What is the physical origin
of the quantum effects on BH entropy? 
Can we calculate the quantum effects in any concrete model? 
Does it make sense to ask about the exact gravitational formula?
Can we compare quantum-gravitational effects to independent 
microscopic calculations of statistical entropy given by the left-hand side of~\eqref{eq:BoltzBH}? 
What are the lessons that can be extracted that apply more generally? 
These are some of the questions that we discuss in Section~\ref{sec:QuantumEntropyBH}. 

\vskip 6pt

\ndt {\bf 
Black hole phases  in the grand-canonical ensemble and AdS/CFT.}
The second strand of our discussion begins
around the same time that supergravity was being invented,
when Gibbons and Hawking~\cite{Gibbons:1976ue} proposed that BHs may be seen as saddle points 
of the Euclidean path integral that computes the Gravitational Path Integral, 
\be\label{eq:Zpathintegral}
    Z \= \int D g_{\mu\nu}\,D \Psi \,\rme^{-\mathcal{S}[g_{\mu\nu},\Psi]}\,.
\ee
Here $\mathcal{S}$ is the Euclidean gravitational action, $g_{\mu\nu}$ is the spacetime metric, $\Psi$ denotes all other 
fields in the theory (these may be gauge fields, gravitini, scalar fields, etc.), and suitable asymptotic boundary conditions are imposed. 

Even keeping aside the mathematical difficulties, the gravitational path integral~\eqref{eq:Zpathintegral}
suffers from a host of problems: the lack of a good physical definition because of non-renormalizability, 
the conformal mode problem~\cite{Gibbons:1978ac},
no manifest relation with Lorentzian physics, etc. 
However, following the Gibbons-Hawking approach, one can try to evaluate~\eqref{eq:Zpathintegral} 
starting from the semiclassical approximation given by a sum over saddles, 
and then including quantum corrections from field fluctuations. 
In the semiclassical limit, the saddle-points are labelled by classical solutions of the theory, so that 
\be\label{eq:saddle_exp_Z}
Z \,\sim\, \rme^{-I_1} + \rme^{-I_2} + \ldots\,,
\ee
where the solutions~$k=1,2,\dots$ have finite Euclidean action~$I_k$ that depend on the external parameters like temperature. 
Upon dialing these parameters, the dominant saddle 
may change, leading to phase transitions. 

As emphasized by Gibbons-Hawking, an advantage of working in Euclidean signature is that BH 
solutions admit a smooth section which caps off at the position of the Lorentzian horizon, 
thus excising the curvature singularity which is relegated in a disconnected region, `beyond the end of space'. 
Then, upon implementing a suitable renormalization  so as to deal with the infinite volume of non-compact space, 
the action of this section of the solution is perfectly well-defined and determines the 
classical contribution of the BH saddle to the partition function. 

Imposing standard, Dirichlet-like boundary conditions in three or more spacetime dimensions leads us to interpret the gravitational path integral \eqref{eq:Zpathintegral} as a grand-canonical partition function, of which BHs are one of the possible thermodynamic phases.
For instance, consider Asymptotically Anti de Sitter (AAdS) space in~$d+1$ dimensions. 
For~$d\ge 2$, BHs can be considered as finite-energy excitations of vacuum AdS$_{d+1}$ space.\footnote{The case~$d=1$ is more subtle as the natural 
boundary conditions correspond to the microcanonical ensemble.} 
In this regime, the on-shell action of any gravitational solution equals the inverse temperature times the free energy in the grand-canonical ensemble 
The free energy allows us to study the structure of phases of the theory, 
and one finds that the dominant phase at low temperature is vacuum AdS space and 
the AdS BH at high temperature. 
Upon dialing the temperature, the theory undergoes the so-called Hawking-Page phase transition between these two phases~\cite{Hawking:1982dh}.

In a beautiful paper in 1998~\cite{Witten:1998zw}, Witten mapped the Hawking-Page phase transition of BHs in AdS space to the deconfinement transition of gauge theory.
The basic context here is the AdS$_{d+1}$/CFT$_d$ holographic correspondence that relates non-perturbative quantum gravity in~$(d+1)$-dimensional AAdS space to a dual conformal field theory (CFT) that lives 
on the~$d$-dimensional conformal boundary of the AdS space.
The master equation of AdS/CFT identifies the gravitational and field theoretical partition functions,
\be \label{eq:Mastereqn}
Z_{\rm grav} (\beta, \mu, \dots ) \= Z_{\, \rm CFT} (\beta, \mu \dots)\,.
\ee
Here~$\beta$ is the inverse temperature of the theory and~$\mu$ denotes chemical potentials in the grand-canonical ensemble. 
The parameters of the gravitational theory and the CFT on the two sides are related by the AdS/CFT dictionary,
where the sources in field theory are identified with 
non-normalizable asymptotic modes in AAdS gravity (see e.g.~the review~\cite{Aharony:1999ti}). 
In this manner the CFT description provides a non-perturbative definition of quantum gravity 
which is also useful in addressing concrete calculations.

We would now like to use the above template in order to answer questions of the following type:  
``Are there other saddles contributing to the path integral in addition to empty space and the BH? What is the dominant phase for a given set of parameters? In particular, what happens in the extremal limit $\beta\to\infty$? Can we tackle these questions using holography?''
These are some of the questions we address in Sections~\ref{sec:grandcanindex} and~\ref{sec:asympt_AdS}.

\vskip 6pt

\ndt {\bf The role of supersymmetry}. 
We are now able to answer some of the questions posed above due to progress in the context of supersymmetric BHs and their microstates. 
In this context, we study supersymmetric versions of the master 
equation~\eqref{eq:Mastereqn}, which take the form of {\it Witten indices}~\cite{Witten:1982df} refined by various chemical potentials. 
We then ask how exactly supersymmetric black holes contribute to the supersymmetric index and what its general saddles are.
Progress has been made in the field theory aspects of the problem as well as in supergravity. 
On the field theory side, different new techniques have been developed recently to analyze the  supersymmetric index of the boundary CFT in great detail, which 
show that the growth of states of the CFT contains the entropy of the supersymmetric BH in AdS space.

\vskip 4pt

Our main focus in this article is to describe the progress on the bulk supergravity side, 
especially for calculating supersymmetric observables.
In particular, we discuss both microcanonical and grand-canonical partition functions 
of the form of refined Witten indices.\footnote{The change of ensemble 
is implemented in the usual way by Laplace transformations. In gravity, the relation between the microcanonical and grand-canonical 
observables appears as the existence of an AdS$_2$ throat inside a asymptotic higher-dimensional AdS space.} 
The index is protected by supersymmetry in the sense that it is invariant under changes of coupling, 
hence we can extrapolate the microscopic result to the gravitational regime and make exact comparisons. 

One point that is brought out in this discussion is
how the microcanonical index, although defined as the difference of number of bosonic and fermionic states with the same quantum numbers, 
actually captures the ground state degeneracy that gives the BH entropy.
Another broader point is that we should take seriously the existence of complex saddles in supergravity. The question of what type of saddles should be considered in the GPI is
an old one. The new aspect here is that the supersymmetric index 
acts as a precise microscopic guide for the choice of gravitational saddles in this case. 

\vskip 4pt

The structure of the rest of the article is as follows. In Section~\ref{sec:QuantumEntropyBH} we discuss the microcanonical version of the 
supersymmetric index and its interpretation as an~AdS$_2$ path integral.
This index, including all-order and non-perturbative quantum corrections, can be calculated 
using the powerful tool of {\it supersymmetric localization}. While this tool has been used successfully in field theory,
its application to supergravity is relatively new, and 
this is what we discuss here.  
In Section~\ref{sec:grandcanindex} we show how to define and calculate, at the 
semi-classical level, the gravitational version of the supersymmetric index.
The saddle-points of the gravitational index are given by a family of complex non-extremal 
supersymmetric solutions labelled by one parameter which formally plays the role of temperature. 
The on-shell action of these solutions is independent of this temperature, and turns out 
to be exactly equal to the action of the leading saddle of the corresponding field theory. We also discuss how one can incorporate higher-derivative corrections in the on-shell action calculation that precisely match the dual field theory prediction.
Further, one finds an infinite number of new saddles also corresponding to a sub-leading exponential growth of states, 
which are given by new orbifolds of the supersymmetric non-extremal saddles of the index. 
We discuss this in Section~\ref{sec:asympt_AdS}.
In Section~\ref{sec:BHsflatspace} we discuss similar non-extremal saddles of the gravitational index in asymptotically flat space.

\section{Quantum entropy of a black hole}\label{sec:QuantumEntropyBH}

In this section we sketch the main physical ideas underlying the concept and calculation of 
exact black hole entropy in supergravity from the microcanonical perspective. 

\vskip 6pt

\ndt {\bf Origin of quantum corrections.}
There are two sources of corrections to the semi-classical Bekenstein-Hawking area law in the full quantum theory.
The first source of corrections stems from the fact that, from the point of view of effective field theory, the two-derivative Einstein-Hilbert action coupled to matter
can receive corrections from local higher-dimensional operators e.g.~terms with higher-derivatives of the 
metric.\footnote{If we have access to the UV description, these terms would be the result of integrating out 
fields and fluctuations above a certain energy scale.}
In the presence of such operators, the Bekenstein-Hawking law does not obey the laws of gravitational 
thermodynamics, and 
it is replaced by a formula known as the Wald formula~\cite{Wald:1993nt,Iyer:1994ys}. This formula is constructed using Noether methods so as to obey the first law
of thermodynamics for any given local effective action. 
The question of what the correct effective action of gravity is at high energies is a difficult one,
which we will not address in any detail here. 
The second source of corrections are quantum fluctuations, i.e.~loops, of light fields in the BH background.
These loop effects for any given effective action 
are sensitive to large-distance properties (and can be surprisingly large), and will be the main focus of the discussion below.

\vskip 6pt

\ndt {\bf Extremal and supersymmetric BHs.} 
As a simple example to illustrate the discussion, 
let us consider four-dimensional theory of the metric and one gauge field governed by the 
Einstein-Maxwell action, 
\be \label{eq:EMaction}
\mathcal{S} \= \frac{1}{16 \pi} \int \diff^4 x\, \sqrt{-g} \, \bigl( R - F_{\mu \nu} F^{\mu \nu} \bigr) \,. 
\ee
This theory admits the Reissner-Nordstrom BH solution, 
with mass~$M$ and electric charge~$Q$.\footnote{Magnetic charges can easily be included in this discussion, but we will not do so here. 
The results including magnetic charges are consistent with electric-magnetic duality.}
The BH has an inner and outer horizon and 
the extremal limit is when the two horizons lie on top of each other, so that~$M=Q$. 
In the near-horizon region of this extremal BH, the metric has the form of~AdS$_2 \times S^2$ 
\be \label{eq:ads2s2}
\diff  s^2 \= Q^2(\diff \eta^2 + \sinh^2 \eta \, \diff \tau^2) + Q^2(\diff \theta^2 + \sin^2 \theta \diff \phi^2) \,,
\ee
with electric field~$F_{\eta \tau} = -\ii Q \sinh \eta$. Here we have presented the metric and gauge fields in Euclidean signature, 
since this is what is used in the path integral calculations below.
This near-horizon configuration is thus completely specified by the charge~$Q$, and it can be checked that it 
is a solution in its own right of the action~\eqref{eq:EMaction}. 
It is easy to read off from the metric~\eqref{eq:ads2s2} that the Bekenstein-Hawking entropy of the extremal BH is given by~$\pi Q^2$. 

The action~\eqref{eq:ads2s2} is, in fact, the bosonic part of the action of pure ungauged~$\CN=2$ supergravity in four dimensions. 
This theory has two gravitini fields which minimally couple to the metric and the gauge field (see e.g.~\cite{Freedman:2012zz} for details). The configuration~\eqref{eq:ads2s2} 
admits eight Killing spinors, 
so that it is a maximally supersymmetric solution of the theory.

\vskip 6pt

\ndt {\bf Attractor mechanism.} 
The main features of the preceding simple discussion carry over to the more 
complicated supersymmetric black hole solutions of supergravity. 
Type IIA string theory compactified on a Calabi-Yau 3-fold is described at low energies by~$\CN=2$ supergravity coupled 
to a number of vector multiplets and hypermultiplets
(see e.g.~\cite{Freedman:2012zz,deWit:1980lyi,Bodner:1990zm} 
for the general structure of~$\CN=2$ supergravity.) 
This theory admits supersymmetric BH solutions which only couple to the metric and the vector multiplets.
The vector multiplets contain 
complex scalar fields that are non-trivially involved in the BH solution. 
This leads to the following tension: on one hand the asymptotic values of the scalar field values (the moduli) can be 
continuously tuned, while on the other hand we expect the BH entropy to be the logarithm of 
an integer as in~\eqref{eq:BoltzBH}, which clearly cannot change continuously.

This tension is resolved by the~\emph{attractor mechanism} which is
one of the very beautiful chapters of the theory of supergravity~\cite{Ferrara:1995ih}. 
What happens is that the supersymmetry of the BH solution leads to a first-order flow equation for all the fields in the theory. The fixed point of this system (the ``attractor'') is the near-horizon configuration, 
where the metric takes the maximally supersymmetric AdS$_2 \times S^2$ with fixed gauge field strengths exactly as in the above simple example. 
Further, in the near-horizon geometry, all the scalar fields become massive and get fixed to constant values determined by the charges. 
In fact, after a rotation in field space~\cite{Sen:2012kpz,Freedman:2012zz},
the near-horizon geometry is effectively described by the metric and one gauge 
field (the graviphoton) governed by the Lagrangian~\eqref{eq:EMaction}. 

The attractor mechanism has been developed further to include the effects 
of higher-derivative terms in the effective action,
and the resulting formula which specializes the Wald formula to supersymmetric BHs 
has been checked to be consistent with the statistical entropy formulas from 
string theory in all cases where one expects an agreement~\cite{LopesCardoso:1998tkj,Sen:2007qy}.

\vskip 6pt

\ndt {\bf Quantum entropy and AdS$_2$/CFT$_1$}. 
The next idea is to promote the above semi-classical set-up to a genuine quantum treatment. 
This needs an imaginative leap in the reasoning, which appeared in the form 
of the quantum entropy conjecture of Sen~\cite{Sen:2008vm}, following the work of~\cite{LopesCardoso:1998tkj,LopesCardoso:2000qm}, and 
the inspiring conjecture of Ooguri-Strominger-Vafa~\cite{Ooguri:2004zv}. 

Sen's quantum entropy conjecture can be presented as follows. 
In string theory, we can build a supersymmetric BH as a bound state of 
intersecting branes extended in the internal compactification directions. 
A natural idea is to consider the decoupling limit at very low energies.
Following a well-established route, one is led to a putative AdS$_2$/CFT$_1$ 
correspondence. 
As we discuss below, there is a deep problem of whether we can genuinely decouple the BH in the quantum theory, but for now we assume this to be the case. 
Now, if the CFT$_1$ has an energy gap between supersymmetric ground states 
and non-supersymmetric excited states, we can decouple the  ground states to obtain a theory with 
Hamiltonian~$H=0$ and hence no 
dynamics.
The partition function of such a 
CFT \footnote{Unfortunately, unlike higher-dimensional CFTs, an explicit controllable example of a CFT$_1$ arising from branes is not known at present. 
However, with some assumptions 
(e.g.~no degrees of freedom in the gravitational theory between the horizon and infinity), 
we can sometimes calculate its supersymmetric index 
using the knowledge of microscopic string theory.} 
would be~$Z_{\text{CFT}_1}= {\rm Tr}_{\mathcal{H}_{\text{CFT}_1}} 1 = \text{dim} (\mathcal{H}_{\text{CFT}_1})$.
Translating this equation to the bulk via the AdS$_2$/CFT$_1$ 
then allows us to rewrite the Boltzmann equation for BHs as follows
\be \label{eq:AdS2CFT1}
Z_{\text{CFT}_1(q,p)} \= \text{dim} (\mathcal{H}_{\text{CFT}_1}(q,p)) \; \equiv \;  
\text{dim} (\mathcal{H}_\text{BH}(q,p)) \=  Z_{\text{AdS}_2}(q,p) \,. 
\ee
Here we have implicitly identified the putative Hilbert space of the BH 
with that of the near-horizon AdS$_2$
or, equivalently, of the dual CFT$_1$. 
Since the BH is taken to be in the microcanonical ensemble, all quantities are computed at fixed charges~$(q,p)$. 
Sometimes the expression in~\eqref{eq:AdS2CFT1} in written
as the exponential of the exact quantum entropy~$\exp \bigl(S_\text{BH}^{\text{qu}}(q,p) \bigr)$. 

The equations in the above paragraph are formal expressions of the ideas of AdS/CFT as applied to the BH entropy problem. 
In order to sharpen these ideas, we recall that the AdS$_2$ path integral has an independent meaning in gravity, at least at the semi-classical level. 
The recent progress in exact quantum BH entropy has arisen from efforts to  push this semi-classical 
formalism to a more precise quantum evaluation of the gravitational path integral. 
The path integral is written as follows,
\be \label{eq:ZAdS2}
\exp \bigl(S_\text{BH}^{\text{qu}}(q,p) \bigr) \=  
Z_{\text{AdS}_2}(q,p) \= 
\int Dg_{\mu\nu} \, D\psi_\mu \, DA_{\mu} \, D\Phi \, 
\rme^{-\cal{S}_\text{sugra}} \, e^{\ii q \oint A}  \,.
\ee
The integral runs over all fluctuations of the gravitational fields of the theory
obeying the asymptotic~AdS$_2 \times S^2$ boundary conditions of the semi-classical near-horizon configuration. 
The Wilson line operator, corresponding to all the electric charges~$q$
carried by the BH, is inserted to impose the microcanonical boundary conditions for the electric charges, as discussed above. 
The magnetic charges~$p$, which are topological in nature, are fixed in terms of integrals over~$S^2$. 
The action~$S_\text{sugra}$ is a function of the metric~$g_{\mu \nu}$, 
the gravitini~$\psi_\mu$, the gauge fields~$A_\mu$, and bosonic and fermionic matter fields~$\Phi$.
It has a bulk piece given by the local action of supergravity, as well as a boundary piece  generalizing the Gibbons-Hawking-York action, which fixes the boundary conditions of all the fields.
The boundary action is arranged such that the saddle-point value of the path integral gives exactly the Bekenstein-Hawking-Wald entropy corresponding to
the local effective action of supergravity~\cite{Sen:2008vm}. 

\vskip 2pt

How to make sense of and calculate the path integral~\eqref{eq:ZAdS2} in the quantum regime is a challenging problem.
A priori, the gravitational path integral suffers from the usual problems of non-renormalizability, unboundedness,
etc. However, the fact that there is an independent meaning of the path integral through the Boltzmann or AdS/CFT equation acts as a guiding principle. 
The leading correction to the Bekenstein-Hawking area law comes from the 1-loop term. 
The entropy including this term is of the form~$\frac14 A_\text{H}+ c_\text{log} \log A_\text{H}$. 
This effect and, in particular, the coefficient~$c_\text{log}$, was calculated by Sen and collaborators~\cite{Banerjee:2010qc,Banerjee:2011jp} (see the review~\cite{Sen:2012kpz}). 
Remarkably, $c_\text{log}$ only depends on field content of the two-derivative gravitational theory, and agrees with 
the microscopic answer given by string theory in all cases that can be calculated. 
In fact, these calculations can also be 
extended to non-supersymmetric BHs including the Schwarzschild BH, and the resulting coefficients  constitute a non-trivial low-
energy constraint to any UV theory that can independently calculate the statistical entropy of a BH.

\vskip 6pt

\ndt {\bf Quantum entropy from localization of supergravity}. 
The evaluation of the quantum entropy of the BH was taken much further in a series of works initiated 
in~\cite{Dabholkar:2010uh,Dabholkar:2011ec}, by using the idea of supersymmetric localization in supergravity. 
Supersymmetric localization is a powerful  technique that calculates 
observables protected by supersymmetry~\cite{Witten:1988ze}. 
It was originally developed for  supersymmetric quantum field theories and has led to remarkable progress in 
calculating many types of supersymmetric observables~\cite{Nekrasov:2002qd,Pestun:2007rz} (see the review~\cite{Pestun:2016zxk}). 

Let us consider a path integral of the  schematic form~$Z=\int \exp(-\cal{S})$, where the action~$S$ and the measure of the integral are invariant under a supercharge~$\mathcal{Q}$ 
The basic idea of localization is to deform the integral to   
\be
Z(\lambda) \= \int \rme^{-{\cal S} - \lambda \mathcal{Q} V} \,, 
\ee
where~$\mathcal{Q} V$ is a~$\mathcal{Q}$-exact term chosen so as to obey~$\mathcal{Q}^2 V=0$. Differentiating the integral in~$\lambda$ brings down a factor of~$\mathcal{Q}V$. 
By considering the action of the supercharge as a differential operator in field space, 
and using that the action and measure are invariant, one can express the integral as a total derivative.
Assuming no boundary terms in field space, one obtains that~$Z'(\lambda) =0$, 
so that the integral of interest~$Z(0)$ equals~$Z(\infty)$.  
As~$\lambda$ becomes larger, the integral is dominated by the critical points of~$\mathcal{Q}V$, called the localization locus.
In the limit~$\lambda \to \infty$ the integral reduces exactly to the semiclassical approximation i.e.~critical points + (one-loop)  determinant of quadratic fluctuations for the~$\mathcal{Q}V$ action. 
In this manner, the integral over the entire field space is reduced to the much smaller subspace of configurations that are invariant under the action of supersymmetry. 
Since the manipulations of the action of supercharge are done for arbitrary field configurations, 
it is important that the supersymmetry algebra closes off-shell for the above logic to hold. 

\vskip 4pt

Unlike for supersymmetric field theory, the initial path integral for supergravity is not well-defined.
The idea of~\cite{Dabholkar:2010uh,Dabholkar:2011ec} is to follow one's nose and obtain a reduced formula  
representing the localized path integral, 
which does make sense and is useful for calculations. 
Before we begin to develop any such idea,  
we need a realization of off-shell supersymmetry transformations for supergravity. 
Fortunately, such a formalism is available for some classes of supergravity, the most important one for this problem being 
$\CN=2$ four-dimensional conformal supergravity. 
This formalism involves auxiliary fields and was developed for completely 
different reasons, namely an elegant  construction of supergravity actions~\cite{deWit:1980lyi} (see the book~\cite{Freedman:2012zz}).

\vskip 2pt

Even with the off-shell supergravity formalism in hand, 
the development of localization in supergravity involves many conceptual as well as technical challenges. 
Below, we briefly mention the various problems and a sketch of the solutions,  
referring the reader to the original references for the full developments. 
\begin{enumerate}
\item The localization formalism needs a rigid (global) supercharge~$\mathcal{Q}$ under which the action and the measure are invariant. 
On the other hand, there is no rigid supercharge to begin with in the theory of supergravity.  
Indeed, the formal presentation of supergravity is a theory in which all symmetries are gauged (local). 
In particular, this includes diffeomorphisms and the supersymmetry transformations.
The intuitive solution to this problem when spacetime has a boundary is clear: 
use the boundary symmetries which act globally on the bulk (just as isometries are the global part of diffeomorphisms). 
This can be regarded as the background field method for gauge symmetries. 
However, implementing the background field method in supergravity formalism is a non-trivial problem, because the gauge algebra of supergravity is \emph{soft} i.e., the structure ``constants'' are not constants (as in a Lie algebra) but rather structure functions of the fields. 
This problem has a nice solution using a certain deformation of the BRST formalism in the presence of an asymptotic boundary~\cite{deWit:2018dix,Jeon:2018kec}. 
This can be viewed as implementing the idea of giving a vacuum expectation value to the ghost for supersymmetry transformations \cite{Baulieu:1988xs,Costello:2016mgj}   
in curved space with an asymptotic boundary when only the asymptotic value of the superghost is fixed. 
\item After obtaining a rigid supercharge, one has to solve the  localization equations to obtain the reduced space of integration. 
This was solved in~\cite{Gupta:2012cy} 
by the method of constructing Killing spinor bilinears and solving the 
resulting bosonic equations in asymptotically AdS$_2 \times S^2$ backgrounds.
The method is an off-shell version of the idea of $G$-structure formalism~\cite{Tod:1983pm,Gauntlett:2002nw}. 
\item The~$\CN=2$ conformal supergravity formalism applies to supersymmetric actions of 
a certain type, governed by the so-called holomorphic prepotential.
This covers the most general two-derivative action of supergravity coupled to an arbitrary number of vector multiplets. 
At higher-derivative level, the action governed by the holomorphic prepotential captures a subset of possible supersymmetric terms ($F$-type terms) 
which can have an arbitrary number of derivatives. 
In addition, there can also be so-called full-superspace invariants ($D$-type terms). 
In Type II string theory compactifications on Calabi-Yau three-folds, the $F$-terms are calculated by the topological string theory on the same CY$_3$~\cite{Antoniadis:1993ze,Bershadsky:1993cx}. 
All known $D$-type terms, on the other hand, can be shown to not contribute to the quantum entropy~\cite{deWit:2010za,Butter:2013lta,Murthy:2013xpa,Butter:2014iwa}, pointing to a non-renormalization theorem. 
\item The calculation of the 1-loop determinants of the deformation ($\mathcal{Q}V$) action is a non-trivial problem. 
These have been calculated by different methods, including a direct evaluation~\cite{David:2018pex,David:2019ocd}, as well as by using more formal index theorems~\cite{Murthy:2015yfa,Jeon:2018kec}. 
While these give results that are mutually consistent and in agreement with microscopic calculations, 
the precise relation between the various approaches is still not completely clear~\cite{Sen:2023dps,GonzalezLezcano:2023uar}. 
\item The precise measure of the supergravity field space in the localized integral is not completely understood. 
The ultralocal measure has been shown to agree with the detailed microscopic result for~$\frac18$-BPS BHs in~$\CN=8$ string theory~\cite{Iliesiu:2022kny}, but the general problem remains open.  
\item Although hypermultiplets and 
spin-$3/2$ multiplets decouple from the classical BH solutions in asymptotic 
flat space, their quantum fluctuations cannot be ignored and, in fact, they do contribute to the 1-loop determinant~\cite{Sen:2012kpz}. 
The hypermultiplets play a more important role in gauged supergravity,
and despite some progress~\cite{Hristov:2019xku}, 
this problem remains open. 
\item In the original localization calculations in supergravity, only smooth supergravity configurations were included in the field space.
In the best understood example of~$\frac18$-BPS BHs in~$\CN=8$ string theory,
this led to a quantum-gravitational result that is exponentially close to the microscopic integer answers. 
The remaining exponentially small corrections were then explained~\cite{Dabholkar:2014ema,Iliesiu:2022kny} as arising from a series of 
orbifolds in string theory of the form
(AdS$_2 \times S^2 \times T^6)/\mathbb{Z}_c$, where~$c=1,2,3,\dots$. 
The question of which orbifolds contribute for a general supersymmetric BH remains open. 
\item The microscopic counting in string theory is really a calculation of the supersymmetric index of states of the type number of bosons minus number of fermions. (More precisely, it is the so-called helicity supertrace in asymptotically flat space, see~\cite{Sen:2009vz,Dabholkar:2010rm}.)
The localization calculation also applies to the supersymmetric index.
On the other hand, the initial physical motivation for the quantum entropy 
involves the degeneracy (number of states) cf.~Eqn.~\eqref{eq:BoltzBH}. 
It was shown in~\cite{Dabholkar:2010rm,Iliesiu:2022kny} that for supersymmetric BHs in asymptotically flat space, all the states of the quantum ensemble are bosonic and hence the degeneracy equals the supersymmetric index. 
\end{enumerate}

\vskip 6pt

\ndt {\bf Quantum decoupling of supersymmetric BH.} 
A final problem concerns the deep issue of decoupling of the supersymmetric BH from the environment that was mentioned above. 
In the quantum entropy calculations, this is reflected by the presence of certain physical zero modes associated to any
gauge field in~AdS$_2$~\cite{Camporesi:1995fb}.
This includes the gauge field for diffeomorphisms (the metric), gauge 
fields carrying electric charges, and gauge fields for supersymmetry (the gravitini).  
At the one-loop level, their effect on the term~$c_\text{log} \, \log A_\text{H}$ can be calculated using the ultralocal measure~\cite{Sen:2012kpz}. 
However, in the exact treatment, it is important to know the full volume of the 
space of zero modes, which appear as an overall multiplicative factor in the path integral. 

The resolution of this problem turns out to have a beautiful connection with recent developments in 
Jackiw-Teitelboim gravity and the Schwarzian theory~\cite{Sachdev:2015efa,Almheiri:2016fws,Maldacena:2016upp,Stanford:2017thb,Moitra:2018jqs,Iliesiu:2020qvm,Heydeman:2020hhw}.
In particular, 
upon turning on a small temperature $T$, 
the action of these modes in four-dimensional supergravity turns out to be precisely governed by the Schwarzian action~\cite{Iliesiu:2022onk}. 
The overall coupling of the action is given by~$T/E_\text{gap}$, where the scale~$E_\text{gap}$ is typically set by a power of the 
extremal entropy. E.g.~for the extremal charged BH in Einstein-Maxwell theory discussed earlier in this section, this scale is~$E_\text{gap}=Q^3$. 
We can then use the exact calculation of the path integral of the Schwarzian theory~\cite{Stanford:2017thb} to calculate the volume of field space of the nearly-zero modes.

The result, and the meaning of the scale $E_\text{gap}$, depends crucially on whether the BH is supersymmetric or not. 
For extremal non-supersymmetric BHs, the partition function is multiplied by a factor of~$T^{3/2}$. 
This means that as~$T \to 0$ these BHs are not decoupled from the environment. 
The approximately~$\exp(S^{\text{class}}_\text{BH})$ states get spread over a scale set by~$E_\text{gap}$ in the full quantum theory.
This resolves an old puzzle about the breakdown of the adiabatic approximation of BHs at very low temperatures~\cite{Preskill:1991tb},
and shows that indeed the physics is consistent with expectations from third ``law'' of thermodynamics, i.e.~the absence of 
a large zero-temperature degeneracy in the absence of symmetries that protects this degeneracy~\cite{Iliesiu:2020qvm}. 

On the other hand, for supersymmetric BHs, $E_\text{gap}$ measures the non-zero energy gap between the degenerate ground states and the first excited state, and the result of the integral over the super-Schwarzian modes is a finite 
number, which agrees with the microscopic formulas from string theory~\cite{Iliesiu:2022kny}.

\vskip 6pt

\ndt {\bf Calculation of $\text{dim}(\mathcal{H}_\text{BH})$ from supergravity.}
From the points discussed above, the dimension of the BH Hilbert space is a supersymmetric index, which can be 
calculated in microscopic string theory by counting the number of states for a given charge. 
This sets a target for the supergravity calculation. 
Now we give a brief snapshot of how the supergravity calculation agrees with microscopic answer in the best-understood example of~$\frac18$-BPS BHs in $\CN=8$ string theory. 
Interested readers can access more details in the original paper~\cite{Iliesiu:2022kny}
and the review article~\cite{Murthy:2023mbc}. 

$\CN=8$ string theory enjoys a large U-Duality group $E_{7,7}(\mathbb{Z})$.  
The properties of the BH as well as the microstate degeneracies are functions of a 
duality-invariant quartic in the charges which we call~$N$. In particular, the horizon area is given by~$4\pi \sqrt{N}$. 
The microscopic index~$C(N)$ is known since the work of~\cite{Maldacena:1999bp,Sen:2009vz}, and its generating function is 
an example of a special class of functions called Jacobi forms, to which we can apply methods of analytic number theory. 
The powerful modular symmetries of this function allows us to obtain the following convergent analytic formula, called the Hardy-Ramanujan-Rademacher expansions, for 
the coefficients~$C(N)$: 
\be\label{eq:RadexpC} 
 C(N) \=   2{\pi} \, \Bigl( \frac{\pi}{2} \Bigr)^{7/2} \, \sum_{c=1}^\infty 
  c^{-9/2} \, K_{c}(N) \; \widetilde I_{7/2} \Bigl(\frac{\pi \sqrt{N}}{c} \Bigr)  \,.
\ee
Here~$\widetilde{I}_{\rho}$ is a modification of the standard $I$-Bessel function of index~$\rho$, and~$K_c$ is a certain number-theoretic function called the Kloosterman sum
which is essentially a sum over phases encoding very little entropy. 

Recalling that the leading asymptotic expansion of the~$I$-Bessel function for large values of argument is given by the 
exponential, we see that the successive terms in~$c$ are exponentially suppressed.
The asymptotic expansion of the leading~$c=1$ term gives gives $\pi \sqrt{N} - 2 \log N$,
which confirms the Bekenstein-Hawking area law as well as the leading logarithmic corrections of~\cite{Sen:2012kpz}. 
The result of the localization calculation in supergravity is~$\widetilde I_{7/2} \bigl(\pi \sqrt{N} \bigr)$, i.e.~precisely the full~$c=1$ term~\cite{Dabholkar:2010uh,Iliesiu:2022kny}. 

The sum over~$c$ is explained by the ~$\mathbb{Z}_c$ orbifolds 
of AdS$_2$ mentioned above. 
The Kloosterman sum is explained by the path integral over certain topological degrees of freedom (Chern-Simons) theory that appear in the orbifolds~\cite{Dabholkar:2014ema}.
The full result of the localization of non-zero modes around the~$\mathbb{Z}_c$ orbifold turns out to be exactly the~$c^\text{th}$ term of~\eqref{eq:RadexpC}, but with a factor of~$1/c$ missing. 
The non-trivial effect of the super-Schwarzian modes around the~$\mathbb{Z}_c$ orbifold 
results in precisely the factor of~$1/c$ compared to the leading saddle~\cite{Iliesiu:2022kny}. 
Thus, upon putting these volumes of the super-Schwarzian modes with the rest of the localized path integral, one obtains a precise agreement 
with the integer degeneracies calculated from microscopic string theory.

The remarkable final result is that we now have a prototype model of quantum 
BHs, namely $\frac18$-BPS BHs in~$\CN=8$ string theory, 
in which we can calculate the precise integer dimension of the BH Hilbert space in the gravitational variables.

\section{The supergravity index in the grand-canonical ensemble}\label{sec:grandcanindex}

\subsection{The gravitational path integral as a sum over saddles} 

We have seen in the introduction that the gravitational partition function~\eqref{eq:Zpathintegral} can be expanded in the semiclassical approximation as a sum over saddles given by solutions to the classical equations of motion. In the following we elaborate on the thermodynamical interpretation of this expansion, as a preparation to the discussion of the supersymmetric case, which will be our main focus next.

The partition function  \eqref{eq:Zpathintegral} is a function of the boundary conditions, hence we discuss these first.
 We are interested in the case where the asymptotic symmetries are such that one can define some conserved global charges measured at infinity. 
 In particular, 
 we assume that these conserved quantities include
 the energy $E$ associated with time translations, a set of U(1) angular momenta~$J_i$, $i=1,2,\dots$ (their number depends on the spacetime dimension), and electric charges~$Q_I$, $I=1,2,\ldots$, associated with gauge symmetries.\footnote{Angular momenta and electric charges are related by dimensional reduction: when these $d$-dimensional gravity models are uplifted into higher-dimensional M- or string theory, the electric charges can be geometrized into angular momenta in the internal directions; conversely, in a reduced two-dimensional theory such as the one defining the AdS$_2$ path integral in the previous section, all angular momenta become electric charges.}$^{,}$\footnote{In four dimensions, we could also consider magnetic charges. These play an important role in AdS black holes where supersymmetry is realized via the topological twist, for instance~\cite{Romans:1991nq,Cacciatori:2009iz}. Although we will not explicitly discuss it here, one can see that the arguments below extend to the case with magnetic charges too.} We also compactify the Euclidean time $\tau=\ii t$ on a circle $S^1$ of length $\beta$, the latter having a standard interpretation as inverse temperature. Then one should specify how the space coordinates and the dynamical fields are identified when going one full time around the Euclidean time circle. Indeed, the identifications can involve a  
 transformation generated by the angular momenta and electric charges, and one should specify the corresponding weights, namely the angular velocities $\Omega^i$ and electrostatic potentials $\Phi^I$, that we will collectively denote as \emph{chemical potentials}.  
Denoting by $\phi^i$ the angular coordinate advanced by $J_i$, 
a revolution in Euclidean time gives the coordinate identifications
 \be
 (\tau+\beta,\,\phi^i -\ii \beta\Omega^i) \;\sim\; (\tau,\,\phi^i)\,,
 \ee
 while any field $\Psi$ is identified as
 \be\label{eq:field_identification}
 \Psi(\tau +\beta,\,\phi^i -\ii \beta \Omega^i ) \= (-1)^\FF \, \rme^{\beta\Phi^I q_I} \, \Psi(\tau,\,\phi^i)\,,
 \ee
 where $\FF$ is the fermion number equal to 0 on bosons and 1 on fermions, 
 and $q_I$ is the charge of the field under $Q_I$ (here we are omitting the other coordinates, which do not get transformed).\footnote{Sometimes it is useful to consider an equivalent picture, obtained by performing the coordinate transformation $\phi^i =\hat \phi^i -\ii \Omega^i \hat\tau$, $\, \tau = \hat\tau$ together with large gauge transformations along the Euclidean time circle, such that one retrieves simpler 
 identifications for the coordinates and the fields,  $ \Psi(\hat\tau+\beta,\,\hat\phi^i ) \,=\, (-1)^\FF \, \Psi(\hat\tau,\,\hat\phi^i)$, but introduces 
 background holonomies around the Euclidean time circle
 at infinity for the metric and the gauge field components that couple to the angular momenta and electric charges.
 } 
 Upon then 
choosing suitable (Dirichlet-type) asymptotic boundary conditions for the metric and the other fields, $Z$ has a statistical interpretation as a {\it grand-canonical partition function} depending on the inverse temperature and chemical potentials,
\be\label{eq:thermal_Z}
Z(\beta,\Omega^i,\Phi^I) \= {\rm Tr} \, \rme^{-\beta(E - \Omega^i J_i- \Phi^I Q_I)}\,.
\ee

In gravity the trace is not really well-defined since we do not know the Hilbert space and the nature of the microstates
making up the statistical ensemble. 
However, when the spacetime is asymptotically AdS and there is a weakly-coupled CFT dual, it can be defined as a trace over the Hilbert space of the CFT in radial quantization. 
Below, we find it convenient to 
reason in terms of operators acting on states and manipulate the trace~\eqref{eq:thermal_Z}, 
even though the gravitational Hilbert space is not known, keeping in mind that these manipulations should be translated into the path integral language. 

It is not hard to see the connection of the partition function~\eqref{eq:thermal_Z} in the thermodynamic regime with 
the Euclidean on-shell action given in~\eqref{eq:saddle_exp_Z}. 
Writing the partition function as a sum over the allowed values of the charges~$(E,J,Q)$ 
(we omit the indices for simplicity), and writing the degeneracy of states with assigned charges as $d(E,J,Q) = \rme^{S(E,J,Q)}$, where $S$ is the entropy, we arrive at the expression 
$Z(\beta,\Omega,\Phi) = \sum_{E,J,Q} \rme^{S(E,J,Q) -\beta(E - \Omega J - \Phi Q)}$.
In the saddle point approximation this can be written as 
\be
Z(\beta,\Omega,\Phi) \; \sim \; \rme^{S(E,J,Q) -\beta(E - \Omega J - \Phi Q) } \,,
\ee
with the saddle point values of $(E,J,Q)$ being fixed by extremization equations which, in the continuum limit, are encoded in the first law of thermodynamics,
\be\label{firstlaw}
\diff E \= \beta^{-1} \diff S + \Omega\, \diff J + \Phi\, \diff Q\, .
\ee
This relates the changes in the thermodynamic quantities when a solution is varied into an infinitesimally close one of the same type.
Comparing this saddle with $Z\sim\rme^{-I}$ we deduce a formula known as the {\it quantum statistical relation}~\cite{Gibbons:1976ue},
\be\label{eq:QSR}
I \= -S + \beta(E-\Omega J-\Phi Q)\,.
\ee
 This shows that the gravitational on-shell action is related to the entropy by a Legendre transformation, with $I/\beta$ playing the role of the Gibbs free energy. Relation~\eqref{eq:QSR} is satisfied by rotating, charged BH solutions with entropy $S$ and can be proven under general assumptions using Wald's method~\cite{Wald:1993nt,Iyer:1994ys}.

\subsection{The gravitational supersymmetric index}

We now focus on
the case where the gravitational theory is supersymmetric.
We choose a complex supercharge, $\mathcal{Q}$, and consider its transformation under the bosonic charges,
\be\label{comm_chargesupercharge}
[J_i,\mathcal{Q}] \=  s_i\mathcal{Q}\,,\qquad [Q_I,\mathcal{Q}] \= - r_I\mathcal{Q}\,.
\ee
 We define the angular momenta so that each $J_i$ rotates an orthogonal two-plane, which implies $s_i = \pm\frac12$. Moreover, if $Q_I$ is a canonically normalized continuous R-charge then $r_I = 1$, while if $Q_I$ is a flavour charge then $r_I=0$.
 The superalgebra also contains the anticommutator
\be\label{superalgebra}
\{\mathcal{Q},{\mathcal{Q}^\dagger}\} \= E - \Omega^{i*} J_i - \Phi^{I*} Q_I\,,
\ee
where $\Omega^{i*}$, $\Phi^{I*}$ are fixed coefficients that may possibly be vanishing (the reason for the notation  will become clear in a moment).
We use~\eqref{superalgebra} to trade the energy $E$ appearing in \eqref{eq:thermal_Z} for the supersymmetric Hamiltonian given by the anticommutator of the supercharge with its conjugate. 
Then the partition function can be written as
\be\label{eq:Z_almostsusy}
Z(\beta,\omega^i,\varphi^I) \= {\rm Tr} \, \rme^{-\beta \{\mathcal{Q},\mathcal{Q}^\dagger\}  +\omega^i J_i +\varphi^I Q_I}\,,
\ee
where we introduced the redefined chemical potentials~\cite{Silva:2006xv}
\be\label{eq:redefined_chem_pot}
\omega^i \= \beta\,(\Omega^I - \Omega^{I*})\,,\qquad \varphi^I \= \beta\,(\Phi^I - \Phi^{I*})\,.
\ee

So far we have just made a change of variables, and the partition function is equivalent to the thermal partition function defined at the beginning. 
We now ask if one can tune the chemical potentials so that the partition function {\it only receives contributions from supersymmetric states} (or, in the path integral representation, from supersymmetric field configurations), as  
in this regime it would be under much better quantitative control. 
The condition for this to happen is that $Z$ takes the form of a refined Witten index~\cite{Witten:1982df},
\be\label{eq:Z_Wittenindex}
Z \= {\rm Tr} \, (-1)^\FF \rme^{-\beta \{\mathcal{Q},\mathcal{Q}^\dagger\} }\, \rme^{(\ldots)}\,,
\ee
where $\FF$ is again the fermion number, and the dots denote combinations of the charges $J_i,Q_I$ that commute with the supercharge. 

Upon comparing \eqref{eq:Z_almostsusy} and \eqref{eq:Z_Wittenindex}, we deduce that the condition to be imposed is
\be
\rme^{\omega^i J_i +\varphi^I Q_I}\mathcal{Q} \= - \mathcal{Q} \ \rme^{\omega^i J_i +\varphi^I Q_I} \,,
\ee
which, recalling the commutation relations~\eqref{comm_chargesupercharge}, is equivalent to 
\be\label{eq:susyconstraint}
s_i\omega^i - r_I\varphi^I  \= \pi  \ii n\,,\qquad n\in\mathbb{Z}_{\rm odd}\,,
\ee
where $\mathbb{Z}_{\rm odd}$ denotes the odd integers.
We can use the constraint~\eqref{eq:susyconstraint} to fix one of the chemical potentials (which one depends on the specific setup under study, see the examples below). 
 Note that the different odd $n$'s in~\eqref{eq:susyconstraint} really give the same boundary condition. Indeed, each $J_i$ that rotates a two-plane only takes integer or half-integer values (in particular, $s_i=1/2$), hence the corresponding $\omega^i$ in the trace \eqref{eq:Z_almostsusy} is only defined modulo shifts by $4\pi \ii $, which can be used to shift $n$ by multiples of 2. In the path integral picture, the periodicity of $\omega^i$ is seen by noticing that every field is either periodic (if bosonic) or antiperiodic (if fermionic) when going one full time around the $2\pi$-periodic orbits of the Killing vector associated with $J_i$, hence going twice around the same orbit 
 is equivalent to the identity transformation.
The equivalence of~\eqref{eq:Z_Wittenindex} with the thermal-type~\eqref{eq:Z_almostsusy} under the constraint~\eqref{eq:susyconstraint} gives us a representation of the 
Witten index which helps to obtain bosonic gravitational saddle-points.

\vskip 2pt

The independence of the Hamiltonian trace~\eqref{eq:Z_Wittenindex} on~$\beta$
follows from the well-known pairing argument of~\cite{Witten:1982df} which shows that the 
the trace only receives contributions from supersymmetric states annihilated by $\mathcal{Q}$ 
and~$\mathcal{Q}^\dagger$.
These facts, however, are less obvious in the gravitational path integral representation. 
A formalism to define the quantum path integral for the gravitational index was given in~\cite{deWit:2018dix},  using a deformed version of BRST transformations. 
This formalism makes it manifest that 
only supersymmetric solutions, i.e.~those that admit Killing spinors, contribute to it. 
This can also be checked explicitly by 
studying one-loop corrections from the field fluctuations around a non-supersymmetric configuration 
and showing that the resulting  fermionic zero-modes arising from the broken supersymmetry 
kills the whole contribution of that saddle to the path integral, see e.g.~\cite{Iliesiu:2021are}.

\vskip 2pt

Now we return to the discussion of saddles and, in particular, black holes. We begin by asking whether supersymmetric BHs contribute to the gravitational index defined above. 
The Gibbons-Hawking ideas about generic non-supersymmetric BHs summarized in the introduction, as well as the agreement of the supersymmetric BH entropy with the microscopic indices in string theory,  
suggest that this would be the case. 
However, supersymmetric BHs have peculiar features which make the question subtle.
Firstly, the gravitational chemical potentials 
$\Omega^i,\Phi^I$ 
are frozen in the supersymmetric case to precisely the fixed 
values $\Omega^{i*}$, $\Phi^{I*}$ appearing in the superalgebra \eqref{superalgebra}.
As such, this would say that there is no good thermodynamic variational principle.  
Secondly, the supersymmetric index in the microscopic theory is defined for arbitrary~$\beta$. 
Indeed, a non-zero~$\beta$ is needed for convergence of the index trace~\eqref{eq:Z_Wittenindex}, and supersymmetry then implies that the index is independent of~$\beta$ as mentioned above. 
On the other hand, since supersymmetric BHs 
are also extremal with~$\beta \to \infty$, they apparently do not satisfy the correct boundary conditions for the index in general.
Finally, $\beta \to \infty$ implies that the near-horizon region is infinitely far away from the asymptotic region and the action is IR divergent,
thus making their contribution to the path integral ill-defined
without the introduction of an appropriate regulator. 

As it turns out, the solution to the third (technical-sounding) problem helps to unravel the first two problems. 
Notice that the redefined chemical potentials in \eqref{eq:redefined_chem_pot} take the indefinite form $\infty \cdot 0$ at extremality, which has a chance to yield a finite non-trivial value if a suitable limiting procedure starting from finite $\beta$ is implemented. 
The idea is thus to study supersymmetric saddles of the gravitational index at finite $\beta$, and take the extremal limit at the end. 

Such a family of saddles for the supersymmetric index was identified 
 in~\cite{Cabo-Bizet:2018ehj}. 
These saddles are supersymmetric solutions to the equations of motion with asymptotic boundary conditions precisely those set
by the supergravity index at finite~$\beta$ as defined above. 
As~$\beta \to \infty$ we recover the Euclidean continuation of the supersymmetric BH. 
Considering the metric as a complex solution, 
we can take a middle-dimensional slice such that the solution includes the topology of a cigar for any finite~$\beta$. 
In this sense they are non-extremal saddles to the gravitational index. 
However, at finite $\beta$ there is no slice which makes the metric completely real. Further, chemical potentials, angular momenta, charges and entropy of the solution are generically complex. 
On the other hand, they have good properties in the grand-canonical ensemble. 
In particular, the on-shell action of these saddles is a holomorphic function of the chemical potentials independent of~$\beta$, and is related to the supersymmetric BH entropy in a way to be discussed below.

The work~\cite{Cabo-Bizet:2018ehj} led to a family of complex  supersymmetric saddles which regulates the supersymmetric BH by a finite temperature. 
It is important to note that their Lorentzian continuations have closed timelike curves and naked singularities, so we would usually remove such solutions from any discussion of the Lorentzian theory. 
The above features, on the other hand, appear very generally in supersymmetric BHs and suggest that such complex saddles have a non-trivial role to play in the Euclidean path integral.
 Indeed, the Euclidean on-shell action 
 gives the correct saddle-point value to the index. 
Further, in cases where they have been computed, the one-loop determinant~\cite{Anupam:2023yns} and non-perturbative corrections~\cite{Aharony:2021zkr} are also consistent with the microscopic theory. 
However, their precise Lorentzian 
interpretation remains an important open question.
We will illustrate further the specific setup of~\cite{Cabo-Bizet:2018ehj} in Section~\ref{sec:aAdS_blackholes}. Here, we continue discussing the general picture which has turned out to be valid for all known supersymmetric black hole solutions.

Whenever the supersymmetric non-extremal solutions outlined above can be defined, it is expected that 
their on-shell action gives a sensible saddle-point contribution to the supergravity index. 
Indeed, the 
quantum statistical relation \eqref{eq:QSR} specialized to supersymmetric solutions 
 and expressed using the redefined chemical potentials \eqref{eq:redefined_chem_pot} as independent variables, takes the form 
\be\label{eq:QSRsusy}
I(\omega,\varphi) \= -S(J,Q) -\omega^i J_i-\varphi^I Q_I\,,
\ee
so we expect it to be finite provided  $\omega^i,\varphi^I$ are finite. In all the cases that have been studied, the on-shell action turns out to be a homogeneous function of degree one in its variables $\omega^i$, $\varphi^I$.
Also, since on supersymmetric configurations the right hand side of \eqref{superalgebra} vanishes, we infer that the first law of thermodynamics~\eqref{firstlaw} reduces to
\be\label{firstlawsusy}
\diff S + \omega^i \diff J_i + \varphi^I \diff Q_I \= 0\ ,
\ee
providing the relation between the chemical potentials $\omega^i$, $\varphi^I$ (subject to the constraint) and the charges $J_i$, $Q_I$. 
Since \eqref{eq:QSRsusy}, \eqref{firstlawsusy} do not depend on $\beta$, we can now take the extremal limit $\beta\to\infty$ and associate well-definite, non-trivial thermodynamic quantities to supersymmetric extremal black holes.
The entropy can be seen as the logarithm of the microcanonical partition function, and it is a function of the conserved charges, $S=S(J,Q)$. Eqns.~\eqref{eq:QSRsusy}, \eqref{firstlawsusy} tell us that this is related to $I(\omega,\varphi)$ by a Legendre transformation subject to the linear constraint~\eqref{eq:susyconstraint}. Namely, the entropy is given by the extremization condition
\be\label{SfromLegTransf0}
S(J,Q) \= {\rm ext}_{\{\omega^i,\varphi^I,\Lambda\}} \left[ -I(\omega,\varphi) -\omega^i J_i  -\varphi^I Q_I -2\Lambda(
s_i\omega^i - r_I\varphi^I  - \pi  \ii n
)\right],
\ee
where the Lagrange multiplier $\Lambda$ implements the constraint. Notice that if  there was no constraint, or if $n=0$, then the Legendre transform would be identically zero due to the degree-one homogeneity of $I$. Instead, one obtains
\be\label{SfromLegTransf}
S \= {\rm ext} \left( 2\pi \ii n\Lambda\right)\,,
\ee
where $\Lambda$ is determined by the extremization equations stemming from \eqref{SfromLegTransf0}, and the ``ext'' symbol persisting in~\eqref{SfromLegTransf} indicates that one should pick the solution for $\Lambda$ such that $2\pi \ii n \Lambda$ has the largest real part.
In the cases that have been studied, the equation for $\Lambda$ is a polynomial equation whose coefficients are real functions of the charges $J_i,Q_I$. Assuming the charges are real,  the solutions for $\Lambda$ are either real, or come in complex conjugate pairs. In order for $S$ to have a non-vanishing real part, we should pick the complex conjugate pairs of $\Lambda$. Hence we obtain a complex expression for~$S$.\footnote{Note that complex conjugate roots lead to a sum over complex conjugate saddles. Indeed, given some odd $n=n_0$, pick the root $\Lambda =\Lambda_0$ that yields the largest real part for $S = 2\pi i n_0\Lambda$. Then for $n=-n_0$, the complex conjugate root $\Lambda =\overline\Lambda_0$ gives $S = \overline S_0 = -2\pi i n_0 \overline\Lambda_0$, which has the same real part. Since, as we have argued, $n_0$ and $-n_0$ correspond to the same boundary condition due to the periodicity of the chemical potentials, it follows that we should sum over these two saddles in the microcanonical partition function, obtaining $Z_{\rm micro} \sim \rme^{S_0} +  \rme^{\overline S_0}= \rme^{{\rm Re}\,S_0}\,2\cos({\rm Im \,S_0}) $. 
(This does not hold if the theory admits a discrete symmetry relating the two configurations, in which case they should not be regarded as distinct saddles).}
Imposing the additional condition  ${\rm Im}\,S = 0$ yields a constraint on the charges that turns out to be precisely the one satisfied in the supersymmetric extremal BH solutions, 
which admit a continuation to a well-definite Lorentzian solution.

  We have seen that the constraint~\eqref{eq:susyconstraint} on the chemical potentials is crucial in order to ensure that the partition function takes the form of an index. We can clarify its  meaning in terms of supergravity field identifications  and argue that it
  is naturally realized in supersymmetric BHs, indicating that the latter do indeed contribute to the index.
The constraint corresponds to imposing the correct boundary condition for the Killing spinor parameter $\zeta$ (as well as the gravitino) 
 when it is transported around the orbits of the Killing vector that generates the horizon. These orbits are contractible at finite $\beta$, so the only allowed spin structure  
 is the antiperiodic one. Since the generator of the horizon has the form $\xi = \partial_t + \Omega^i \partial_{\phi^i}$, we should check that precisely the identification \eqref{eq:field_identification} is respected, including the minus sign which stems from $(-1)^\FF$ and gives the antiperiodic structure. Evaluating \eqref{eq:field_identification} for the spinor $\zeta$, 
  one finds the condition $\rme^{s_i\omega^i - r_I\varphi^I} = -1$,
 which is precisely the statement of the constraint. Hence the condition~\eqref{eq:susyconstraint} which follows from supersymmetry is also the one that is compatible with the cigar-like topology associated with a finite-$\beta$ Euclidean horizon in the bulk. In all BH solutions that have been analyzed, the odd integer $n$ in~\eqref{eq:susyconstraint} is fixed to $n=\pm1$.

We now address the complexification of the solution. The imaginary unit appearing in the constraint~\eqref{eq:susyconstraint} requires that some of the thermodynamic variables $\beta$, $\Omega$, $\Phi$ entering in the definition~\eqref{eq:redefined_chem_pot} of $\omega$, $\varphi$ take imaginary (possibly, complex) values, while ordinarily they are real in black hole thermodynamics.
In order to achieve this, in addition to Wick-rotating the Lorentzian time to Euclidean time as $t = -\ii \tau$, one may analytically continue some of the parameters describing a given Lorentzian solution, which also enter in the expressions of $\beta$, $\Omega$, $\Phi$.
Crucially, this must be done in a way consistent with supersymmetry in the bulk, which also boils down to a condition on the parameters.
 Depending on the setup, it may be possible to keep the analytically continued metric real, or not. Specifically, in asymptotically AdS finite-$\beta$ solutions the metric resulting from these manipulations is complex, while in asymptotically flat solutions it is possible to choose a continuation of the parameters such that the metric is real and positive-definite.\footnote{This is because  asymptotically flat supersymmetric solutions are obtained in ungauged supergravity, where the $r_I$ appearing in the constraint \eqref{eq:susyconstraint} vanish. In this case, the constraint can be satisfied by requiring that just the $\Omega^i$ are purely imaginary, which is achieved by analytically continuing the rotational parameters of the solution to purely imaginary values. Combined with the Wick rotation of the time coordinate, this leads to a real metric.} In all cases, the quantity that would be usually identified as the entropy is a complex function of the angular momenta and charges at finite $\beta$.
 Although it is not a priori obvious what the regularity conditions are that must be imposed on a complex solution, an approach that works 
  is to first make sure that the metric admits a real section that is regular in the usual sense, then use this to obtain the expressions for the thermodynamic quantities in terms of the parameters, and only at this stage impose supersymmetry by
 analytically continuing the solution to the complex domain. These supersymmetric solutions may then be interpreted as complex saddles of the gravitational index. When eventually the $\beta\to\infty$ limit is taken, the solution may be continued back to a well-definite Lorentzian solution.

The ideas above turn out to work nicely for all known supersymmetric black holes in three or more dimensions, both in  AdS \cite{Hosseini:2017mds,Hosseini:2018dob,Cabo-Bizet:2018ehj,Choi:2018fdc,Cassani:2019mms,Kantor:2019lfo,Bobev:2019zmz,Bobev:2020pjk,Larsen:2021wnu,BenettiGenolini:2023ucp}
and  flat space~\cite{Iliesiu:2021are,Hristov:2022pmo,H:2023qko,Boruch:2023gfn,Cassani:2024kjn,Boruch:2025qdq}; this includes solutions with conical singularities such as accelerating AdS$_4$ black holes~\cite{Cassani:2021dwa}. In the following we illustrate the story further using the significant example of~AdS$_5$ black holes.

\section{Asymptotically AdS black holes}\label{sec:asympt_AdS}

It is interesting to study the supergravity index  
with asymptotically AdS boundary conditions using the dual superconformal field theory (SCFT). 
In particular, the leading behaviour of the SCFT index in the large-$N$  expansion is expected to match the weakly-coupled gravitational regime described by two-derivative supergravity. 
Recently, a careful study of different supersymmetric indices in the large-$N$ limit has led to a very precise microscopic derivation of the Bekenstein-Hawking entropy of supersymmetric AdS BHs in different spacetime dimensions \cite{Benini:2015eyy,Cabo-Bizet:2018ehj,Choi:2018hmj,Benini:2018ywd}, with a lot of work having been done in the last decade; see~\cite{Zaffaroni:2019dhb} for a review and additional references. Discussing how this is achieved would go beyond the scope of the present article, which is focused on the supergravity side of the story. 
In the following, we illustrate the general ideas outlined above in the relatively simple case of AdS$_5$ BH solutions to minimal five-dimensional gauged supergravity, only briefly mentioning the relation with the dual field theory results, emphasizing how they have guided progress in the gravitational analysis.

\subsection{The universal five-dimensional supergravity black hole}\label{sec:aAdS_blackholes}

We follow~\cite{Cabo-Bizet:2018ehj}, to which we refer for more details.
 The bosonic Lagrangian of minimal five-dimensional gauged supergravity~\cite{Gunaydin:1984ak} reads\footnote{
  A slightly unusual normalization for the gauge field $A$ is chosen so that the electric charge $Q$ is a canonically normalized R-charge, satisfying \eqref{comm_chargesupercharge} with $r=1$. Its value in any asymptotically AdS solution is thus precisely identified with the one-point function of the dual SCFT R-charge, facilitating comparison of results.
 In these conventions, the ungauged supergravity limit $g\to 0$ needs to be taken with care: first one should rescale $A \to g A$ (and therefore $Q\to Q/g$, $\varphi \to g \varphi$), and only then send $g\to 0$.
\label{foot:conventions_gauge} }
\be
\mathcal{L} \= (R+ 12 g^2)\, {*1} - \frac{2}{3g^2} F\wedge{*F} + \frac{8}{27g^3}
      A\wedge F\wedge F \ ,
\ee
where $A$ is an Abelian gauge field with field strength $F={\rm d} A$, while $g>0$ is the cosmological constant parameter.
This theory admits a supersymmetric AdS$_5$ solution with radius $1/g$, and can be uplifted to ten- or eleven-dimensional supergravity in many different ways, in particular it uplifts to type IIB supergravity on $S^5$ or any other Sasaki-Einstein manifold.
From a holographic viewpoint, it describes the universal dynamics of the energy-momentum tensor multiplet of dual four-dimensional $\mathcal{N}=1$ SCFTs at large $N$ (including in particular  $\mathcal{N}=4$ SYM, seen as an $\mathcal{N}=1$ theory).

 The general asymptotically AdS$_5$ black hole solution  was given in~\cite{Chong:2005hr}. It carries two independent angular momenta $J_1,J_2$ associated to rotation along two angular directions $\phi_1,\phi_2$ in AdS$_5$, and one electric charge $Q$, which is a canonically normalized R-charge, in addition to the energy $E$. 
 The supersymmetric and extremal black hole was found in \cite{Gutowski:2004ez,Chong:2005hr}. Choosing the supercharge such that \eqref{comm_chargesupercharge}, \eqref{superalgebra} hold, the energy of any supersymmetric solution is fixed to
\be\label{superalgebra_5d_gauged}
E \= g J_1 + g J_2 + \frac{3}{2}g Q\,,
\ee
while the combinations of the charges that commute with the supercharge (and that can then appear in a supersymmetric partition function) are given by $J_1+\frac{1}{2}Q$ and $J_2+\frac{1}{2}Q$. 
The extremality condition for the solution of~\cite{Chong:2005hr} imposes an additional constraint on top of \eqref{superalgebra_5d_gauged}, so that the charges $J_1,J_2,Q$ satisfy the non-linear relation 
\be\label{nonlinear_rel_BPScharges}
 \left(3 Q  + \frac{\pi}{2Gg^3} \right)\left( 3 {Q}^2 -  \frac{\pi}{Gg^3}(J_1+J_2)\right) \=  Q^3 + \frac{2\pi}{Gg^3}J_1 J_2\ ,
\ee
where $G$ is the five-dimensional Newton's constant.
The Bekenstein-Hawking entropy of the supersymmetric and extremal solution, evaluated as 1/4 the area of the horizon three-sphere, can be written as
\be\label{S_BPS}
S \= \pi \sqrt{3Q^2 -\frac{\pi}{G g^3} \big(J_1+J_2\big)}\ .
\ee  

We then turn to the chemical potentials. Comparing \eqref{superalgebra_5d_gauged} with \eqref{superalgebra}, we deduce that $\Omega_1^* = \Omega_2^* = g$, $\Phi^* = \frac{3}{2}g$. 
On the other hand, the redefined chemical potentials~\eqref{eq:redefined_chem_pot} are found to satisfy the constraint
\be\label{constraint_chempot_5d}
\omega_1+\omega_2 - 2 \varphi \= \pm 2\pi \ii\,
\ee
and remain finite, complex functions of the parameters when the extremal limit is taken.
The supersymmetry condition \eqref{eq:susyconstraint} is then satisfied, with $n=\pm1$. As discussed in Section~\ref{sec:grandcanindex}, the two possible signs in \eqref{constraint_chempot_5d} yield the same boundary condition. However, the two corresponding bulk solutions are distinct (only coinciding in the extremal limit), and hence we should sum over them in the partition function. 
Since the suitable boundary conditions are satisfied, the partition function $Z$ takes the form of a supersymmetric index~\eqref{eq:Z_Wittenindex}. Specifically, it can be written as
\be\label{Z_as_an_index}
Z \= {\rm Tr} \, (-1)^\FF \rme^{-\beta \{\mathcal{Q},\mathcal{Q}^\dagger\} }\, \rme^{(\omega_1\mp 2\pi \ii) (J_1+\frac12 Q) + \omega_2 (J_2+\frac12 Q)}\,,
\ee
where we used the linear constraint~\eqref{constraint_chempot_5d} to eliminate $\varphi$, as well as the spin-statistics identity~$\rme^{2\pi  \ii J_1}=(-1)^\FF$  to explicitly display the fermion number operator.\footnote{Of course we could have equivalently chosen $J_2$ instead of $J_1$ to realize the fermion number $\FF$.
} This expression agrees with the dual SCFT index originally defined in~\cite{Romelsberger:2005eg,Kinney:2005ej}.

It was noticed in~\cite{Hosseini:2017mds} that the expression \eqref{S_BPS} for the supersymmetric extremal entropy is reproduced by the constrained Legendre transform of a simple function of the chemical potentials. 
This function was then shown in~\cite{Cabo-Bizet:2018ehj} to correspond to the renormalized action\footnote{This is the renormalized action computed via background subtraction, which is such that the AdS$_5$ vacuum solution contributes as 1 to the partition function $Z$.} of the supersymmetric non-extremal black hole solution, and reads
 \be\label{Ionshell_AdS5}
I \= \frac{2\pi}{27G g^3}\,\frac{\varphi^3}{\omega_1\omega_2}\,,
\ee
indicating that the solution is a good saddle of the gravitational index for any values of $\beta$ (including the extremal limit $\beta\to \infty$).

Developments initiated in~\cite{Cabo-Bizet:2018ehj,Choi:2018hmj,Benini:2018ywd} showed that the expression~\eqref{Ionshell_AdS5} for the action matches the computation of the supersymmetric index of the dual SCFT at large~$N$. 
The SCFT index is computed by a matrix integral over gauge holonomies, which can be analyzed via different methods. These include the Bethe ansatz approach~\cite{Benini:2018ywd,GonzalezLezcano:2019nca,Lanir:2019abx,Benini:2020gjh,Aharony:2021zkr}, where the index is rewritten as a sum over solutions to a set of transcendental equations similar to the Bethe ansatz equations of integrable systems;
an elliptic deformation of the matrix integral allowing for the large-$N$ saddle point analysis~\cite{Cabo-Bizet:2019eaf,Cabo-Bizet:2020nkr,Cabo-Bizet:2020ewf}, the results of which are consistent with the Bethe ansatz method~\cite{Colombo:2021kbb}; the study of Cardy-like regime (to be discussed below), and other approaches \cite{Copetti:2020dil,Choi:2021rxi,Choi:2023tiq}. Via the Legendre transform \eqref{SfromLegTransf0}, this match has provided a {\it microscopic derivation of the Bekenstein-Hawking entropy of supersymmetric AdS$_5$ black holes via holography}. 

However, the SCFT index contains much more than just information about the classical on-shell action of the black hole solution we have been considering so far.
Firstly, the index has an intricate large-$N$ structure, with different saddles dominating in different regimes of the chemical potentials, which are expected to be matched by different solutions to the supergravity equations of motion. Secondly, even if we just focus on the single specific large-$N$ saddle whose action matches the black hole on-shell action, it is possible to systematically include perturbative as well as non-perturbative corrections to the leading large-$N$ contribution, thus gaining new insight into the quantum black hole entropy. On the gravity side, the perturbative  corrections should correspond to quantum and higher-derivative corrections, and determining how these arise offers an intriguing opportunity to advance our understanding of quantum supergravity. In the following we briefly review these two aspects, starting from the latter.

\subsection{Corrections from higher-derivative supergravity}

One way to conveniently isolate the saddle of the SCFT index corresponding to the supersymmetric black hole is to take a Cardy-like limit before taking the large-$N$ limit. This is defined as a regime where the angular chemical potentials entering in \eqref{constraint_chempot_5d}  are taken small,  $\omega_1, \omega_2\to 0$, implying that the conjugate charges become large.\footnote{Notice that in the SCFT index \eqref{Z_as_an_index}, this regime corresponds to sending the chemical potential for $J_2+\frac12 Q$ to 0, but the chemical potential for $J_1+\frac12 Q$ to $\mp 2\pi i$. Since shifting one of the $\omega$'s by $2\pi i$ is in general not a symmetry of the index, this is different from the more standard Cardy-like regime studied in~\cite{DiPietro:2014bca}, where both chemical potentials for $J_i+\frac12 Q$, $i=1,2$, are taken small.}
It was shown in a number of papers \cite{Choi:2018hmj,Honda:2019cio,ArabiArdehali:2019tdm,Kim:2019yrz,Amariti:2019mgp,Cabo-Bizet:2019osg,
Gadde:2020bov,
GonzalezLezcano:2020yeb,Goldstein:2020yvj,Amariti:2020jyx,Amariti:2021ubd,Cassani:2021fyv,ArabiArdehali:2021nsx}, using different methods and reaching a progressively increasing accuracy, that the index in this regime is controlled by the 't Hooft anomalies of the SCFT conserved global currents.  
For the universal index we have been discussing above, which only uses the R-charge $Q$ in addition to the angular momenta $J_1,J_2$,
and under mild assumptions, the asymptotic 
behaviour in the Cardy-like regime reads 
\be\label{eq:logI}
\log Z \= -I + \ldots\,,
\ee
where
\be\label{cardyresult}
I \= \kRRR\, \frac{  \varphi^3}{6\,\omega_1\omega_2}  -  \kR\, \frac{ \varphi\,(  \omega_1^2 + \omega_2^2-4\pi^2) }{24\,\omega_1\omega_2} \,,
\ee
and $\kRRR={\rm Tr}\,\hat Q^3$, $\kR = {\rm Tr}\,\hat Q$ are the cubic and linear anomaly coefficients for the SCFT R-charge generator $\hat Q$.
The dots in \eqref{eq:logI} indicate that we are omitting exponentially suppressed terms, as well as a logarithmic term $\log |\mathcal{G}|$, where $|\mathcal{G}|$ is the order of a discrete one-form symmetry group $\mathcal{G}$ that the theory may have, which counts the degeneracy of saddles with the same action~\cite{Cassani:2021fyv}. The expression above is valid at finite $N$.

Let us now assume the SCFT is holographic and consider its large-$N$ expansion. At leading order, a standard holographic  dictionary relates the SCFT anomaly coefficients to the gravitational units as
\be\label{holo_dictionary}
\kRRR\=\frac{4\pi}{9Gg^3}+ \ldots\,\qquad \kR \= 0 + \ldots\,,
\ee
where the dots denote subleading terms. It follows that
 \eqref{cardyresult} matches the expression  \eqref{Ionshell_AdS5} for the two-derivative supergravity on-shell action.

The challenge is then to match in supergravity the subleading terms appearing in the large-$N$ expansion of~\eqref{cardyresult}. In the following we summarize how the question has been answered at next-to-leading order, while an all-order match of the formula has not been achieved to date. We will follow~\cite{Cassani:2022lrk}, see also~\cite{Bobev:2021qxx,Bobev:2022bjm} for closely related results.
Restricting to leading order in the corrections, it becomes easy to work out the Legendre transform of~\eqref{cardyresult} according to~\eqref{SfromLegTransf0} and extract the prediction  for the corrected supersymmetric black hole entropy. This can be done by treating $\kR$ as a small perturbation. The corrected entropy reads
\be\label{eq:microcanonicalentropyCFT}
 S \=\pi\sqrt{3Q^2 -8  a \left(J_1 + J_2\right) - 16\,  a \left({a}- c\right) \frac{(J_1-J_2)^2}{Q^2-2\, a\left(J_1 + J_2\right)}}\,\,,
\ee
while the corrected version of the charge relation \eqref{nonlinear_rel_BPScharges} which ensures reality of the entropy becomes
\be
\begin{aligned}\label{corrected_constraint}
& \left[3 Q  + 4\left(2a-c\right) \right]\left[ 3 Q^2 -  8c\, (J_1+J_2)\right]   \\
&\= Q^3 + 16 \left(3c-2a\right)J_1J_2 +\,64a\, (a-c)\frac{(Q+a)(J_1-J_2)^2}{Q^2-2a (J_1+J_2)}\,.
\end{aligned}
\ee
Here, we have found it convenient to use the Weyl anomaly coefficients  $a = \frac{3}{32}(3\kRRR - \kR)$, $c = \frac{1}{32}(9\kRRR - 5 \kR)$, and it is understood that both \eqref{eq:microcanonicalentropyCFT} and \eqref{corrected_constraint} are only valid at linear order in $\kR$. 

 A hint about how to proceed in supergravity comes from the crucial fact that the SCFT only contributes to the formula~\eqref{cardyresult} through its 't Hooft anomaly coefficients. It is thus reasonable to assume that the needed five-dimensional supergravity theory is the one that reproduces these anomalies holographically. It is well-known that anomalies arise as boundary terms by varying suitable Chern-Simons terms in one dimension more, and that in holography this is precisely the mechanism by which gravity reproduces the dual field theory anomalies~\cite{Witten:1998qj}. The idea is thus to start from the Chern-Simons terms that capture the cubic and linear 't Hooft anomalies for the SCFT R-symmetry, and then complete the action by imposing sypersymmetry.
The Chern-Simons terms of interest are two-derivative and four-derivative terms, and read
\be\label{CSterms_intro}
\mathcal{S}_{\rm CS} \=  \frac{1}{24\pi^2} \int \Big( \kRRR\, A \wedge F \wedge F - \frac{1}{8} \, k_R\, A \wedge R_{ab} \wedge R^{ab} \Big)\,,
\ee
where $R_{ab}$ is the Riemann curvature two-form.
We then need to supersymmetrize~\eqref{CSterms_intro},  
 also implementing a gauging of the R-symmetry so that the resulting supergravity theory admits an AdS vacuum.
 To do so, we can rely on the general structure of off-shell supergravity and exploit its known two-derivative and four-derivative invariants~\cite{Hanaki:2006pj,Ozkan:2013nwa,Gold:2023ymc}. Working at linear order in the couplings governing the four-derivative invariants, the auxiliary fields of off-shell supergravity can be eliminated by solving algebraic equations of motion (which instead become dynamical, and thus much harder to solve, in the full non-linear theory). It is also convenient to implement suitable field redefinitions so as to simplify the Lagrangian.

 The result of these manipulations is a supergravity model that can be seen as a supersymmetric effective action for the dual SCFT 't Hooft anomalies.
  We call $\lambda_1,\lambda_2$ the small dimensionless parameters controlling the off-shell higher-derivative invariants we start from. Then the bosonic Lagrangian reads at first order in the corrections,
\begin{align}\label{corrected_action_5d}
\mathcal{L}  &\=  \left( c_0 R+12c_1g^2\right)*1 -\frac{2c_2}{3g^2}  \,F\wedge * F  + \frac{8c_3}{27g^3}\,A\wedge F\wedge F\nn\\[1mm]
&\quad + \frac{\lambda_1}{g^2}\left[\Big(\mathcal{X}_{\rm GB} - \frac{2}{3g^2}C_{\mu\nu\rho\sigma}F^{\mu\nu}F^{\rho\sigma}+\frac{2}{9g^4}F^4 \Big)*1 + \frac{4}{3g} \,A \wedge R_{ab} \wedge R^{ab} \right] ,
\end{align}
where $\mathcal{X}_{\rm GB} \= R_{\mu\nu\rho\sigma}R^{\mu\nu\rho\sigma} - 4R_{\mu\nu}R^{\mu\nu} +R^2$ 
is the Gauss-Bonnet combination of Riemann curvatures, $C_{\mu\nu\rho\sigma}$ is the 5d Weyl tensor, $F^4= F_{\mu\nu}F^{\nu\rho}F_{\rho\sigma}F^{\sigma\mu}$, and the coefficients specifying the corrections to the two-derivative terms
are
\be
c_0 = 1+4 \lambda_2\,,\quad c_{1}=1-10 \lambda_1 +4\lambda_2 \,, \quad {c}_{2}=1+4 \lambda_1 +4 \lambda_2 \,, \quad {c}_{3}=1-12\lambda_1+4\lambda_2\,.
\ee
Notice that while $\lambda_1$ appears both in the two-derivative in the four-derivative terms in \eqref{corrected_action_5d}, $\lambda_2$ just renormalizes the overall Newton's constant in the action.
Comparing with \eqref{CSterms_intro} we obtain the holographic dictionary \eqref{holo_dictionary} 
between the SCFT anomaly coefficients and the supergravity coupling constants 
 at next-to-leading order,
\be
\kRRR \= \frac{4\pi}{9Gg^3} \left( 1 -12\lambda_1 +4 \lambda_2  \right)\,,\qquad
\kR \= -\frac{16\pi}{Gg^3} \, \lambda_1 \,.
\ee

Now one should evaluate the (suitably renormalized) action on the supersymmetric black hole solution. It turns out that at first-order in the corrections, this can be done without needing to solve the equations of motion corrected by the terms proportional to $\lambda_1,\lambda_2$.  
Also the chemical potentials should not be corrected, as in the grand-canonical ensemble they are independent variables. Then an explicit computation shows that the corrected on-shell action precisely matches~\eqref{cardyresult}. The entropy, the angular momenta and the electric charge, instead, receive non-trivial corrections both because the action is modified, and because the solution is corrected. These can be evaluated by differentiating the on-shell action with respect to the chemical potentials. One can also compare with the Wald entropy extracted from the corrected near-horizon solution (which is easier to determine than the full solution thanks to the enhanced symmetry), finding agreement. The explicit expressions can be found in~\cite{Cassani:2022lrk,Cassani:2023vsa, Cano:2024tcr}.

\vskip 4pt

 The higher-derivative supergravity black hole action presented above successfully reproduces the SCFT prediction~\eqref{cardyresult} at next-to-leading order in the large-$N$ expansion, and it would be instructive to improve on this. 
 A strategy to do so may be to work directly in off-shell supergravity, and possibly exploit localization arguments.
 Specifically, it would be nice to prove the conjecture that the only supergravity invariants contributing to the supersymmetric on-shell action are those involving Chern-Simons terms. This would represent a gauged supergravity counterpart of the non-renormalization results reviewed in Section~\ref{sec:QuantumEntropyBH}. Since in minimal five-dimensional supergravity there are no other Chern-Simons terms beyond the two- and four-derivative terms in~\eqref{CSterms_intro}, this would explain why we have been able to match the full functional form of~\eqref{cardyresult} by just studying the Lagrangian \eqref{corrected_action_5d}. Then one should evaluate the supersymmetric Chern-Simons invariants on the black hole field configuration beyond linear order in the corrections. Finally, one should determine the holographic dictionary between the coefficients of the Chern-Simons supergravity terms and the anomaly coefficients $\kRRR$, $\kR$ beyond linear order.
 The last step requires considering a concrete holographic pair, and uplifting the five-dimensional supergravity theory to higher-dimensions so as to identify the mechanisms in the fundamental theory which generate the five-dimensional Chern-Simons terms in the compactification. In particular, for the $\mathcal{N}=1$ quiver gauge theories that are dual to type IIB supergravity on Sasaki-Einstein manifolds, the anomaly coefficients $\kRRR$ and $\kR$ only contain $\mathcal{O}(N^2)$ and $\mathcal{O}(1)$ terms, hence their large-$N$ expansion truncates at next-to-leading order. On the gravity side, the $\mathcal{O}(N^2)$ terms are generated by dimensional reduction of the type IIB supergravity action, while the corrections are expected to be generated by one-loop effects in the KK tower of type IIB supergravity modes on the Sasaki-Einstein  manifold. It would be nice to prove this in detail, as well as to study the extension to different holographic dual pairs, such as those describing M5-branes wrapped on Riemann surfaces.

While here we have just presented the case where the universal R-charge $Q$ is turned on,  extensions to the multi-charge case have also been studied~\cite{Cassani:2024tvk}, where the thermodynamics of solutions to higher-dimensional supergravity coupled to vector multiplets is compared with the Cardy-like limit of the SCFT  index refined by flavour chemical potentials.
We also note that the program of developing a precise holographic match between higher-derivative corrections in supergravity solutions and subleading corrections to field theory free energies has been carried out
 for AdS$_4$/CFT$_3$ dual pairs in~\cite{Bobev:2021oku}.

\subsection{Sum over saddles and non-perturbative contributions}

We have seen how supersymmetric black hole solutions contribute to the gravitational index. A related question is what is the complete set of supersymmetric gravitational solutions that satisfy the same  boundary conditions,  
 and how they contribute to the index. Answering this question would allow to determine the phase space of the supersymmetric ensemble under consideration.
While a full classification is not known, in the following we discuss how certain solutions, related to each other by discrete transformations, in fact contribute to the same gravitational index. These discrete transformations are specific integer shifts of the chemical potentials as well as new orbifolds of the supersymmetric non-extremal solutions. This will demonstrate that the gravitational index receives contributions from multiple competing saddles.
 The discussion will also lead us to consider some non-perturbative contributions to the saddle point action.
We illustrate this again in the context of asymptotically AdS$_5$ solutions, reviewing results of~\cite{Aharony:2021zkr}. Similar features in an asymptotically AdS$_4$ context were discussed in~\cite{BenettiGenolini:2023rkq}. 
 
We start by recalling that certain suitable shifts of the chemical potentials do not modify the boundary conditions.
 This implies that when evaluating the partition function, we should sum over the solutions carrying the shifted chemical potentials, 
  provided they are admissible saddles. 
For instance, consider the shift of the angular velocities given by $\Omega^i\to \Omega^i + \frac{2\pi \ii}{\beta}n_i$, with $n_i \in \mathbb{Z}$ (which corresponds to $\omega^i \to \omega^i+2\pi \ii n_i$).
This gives an extra factor of $(-1)^{(\sum_i n_i)F}$ in the field identifications~\eqref{eq:field_identification}, where again $F$ is the fermion number.
 Similarly, one can shift the electrostatic potentials, although their periodicities depend on the charges of the different fields under the respective symmetry, which is a model-dependent feature requiring information from the microscopic theory.
As a concrete example, let us consider the complex supersymmetric, non-extremal black hole solution to minimal gauged supergravity discussed in Section~\ref{sec:aAdS_blackholes} and uplift it to type IIB supergravity on $S^5$ with $N$ units of RR five-form flux, which is dual to ${\rm SU}(N)$ $\mathcal{N}=4$ SYM theory. Then the five-dimensional electrostatic potential $\Phi$ is the potential for the U(1)$_R$ charge which rotates in the same way the three orthogonal planes in $\mathbb{R}^6 \supset S^5$. The R-charges of all higher-dimensional bosonic fields are in $\frac{2}{3}\mathbb{Z}$, while the R-charges of all fermionic fields are in $\frac{1}{3}(2 \mathbb{Z} +1)$. It follows that a shift $\Phi \to \Phi + \frac{3\pi \ii}{\beta}n_R$, with $n_R \in \mathbb{Z}$  (which corresponds to $\varphi \to \varphi + 3\pi \ii n_R$) implies that the type IIB fields acquire the extra phase $(-1)^{n_RF}$ when going one full time around the Euclidean time circle. Combining this with the shift of the angular velocities, we see that in order to preserve the original identifications of both bosonic and fermionic fields, we just need to impose
\be\label{eq:constraint_integer}
n_1 + n_2 + n_R \,\in\, 2\mathbb{Z}\,.
\ee

We have thus obtained a large family of bulk chemical potentials corresponding to the same boundary conditions.
It follows that the different black hole solutions corresponding to these chemical potentials should be summed over in the gravitational path integral computing the index, each contributing with its own on-shell action~\eqref{Ionshell_AdS5}.
An important point is that, for the solution under consideration to admit a Killing spinor in the bulk, the constraint~\eqref{constraint_chempot_5d} also needs to be satisfied, which restricts the choice of integers to be summed over in the partition function. 

However, the authors of~\cite{Aharony:2021zkr} argued that  
not all supersymmetric solutions related through the shifts above should be included in 
  the gravitational partition function: the following admissibility criterion was proposed, based on brane stability. 
 Embed an instantonic probe brane in a supergravity solution with bulk action $I$, and call $I_{\rm brane}$ the brane action in this background. This yields a contribution to the gravitational partition function of the form $\rme^{-I} \rme^{\ii I_{\rm brane}}\,,$
which should be added to the pure supergravity contribution, $\rme^{-I}$.\footnote{The factor of $\ii$ in front of the brane action is not cancelled out by the Wick-rotation $t=-\ii \tau$ because the brane does not extend along the time direction.}
 The new contribution is exponentially suppressed if
\be\label{stability_criterion}
{\rm Im}(I_{\rm brane}) \,>\, 0\,,
\ee
in which case it can be seen as a non-perturbative correction to the saddle-point action:
\be\label{eq:I_corrected}
I_{\rm corrected}\,\simeq\, I - \rme^{\ii I_{\rm brane}}\,.
\ee
If any of the possible brane embeddings fails to satisfy condition \eqref{stability_criterion}, then  including an increasing number of such branes would lead to a divergent series with large terms, which signals an instability of the supergravity solution towards brane condensation, presumably driving the system to a different classical solution.
 Hence, according to the proposed criterion the unstable supergravity solution should not be included in the sum over saddles when evaluating the partition function in the semiclassical expansion.

In the type IIB setup we have been considering, the relevant branes are D3-branes, since the other stringy objects carry a dependence on the continuous 't Hooft coupling and are therefore expected not to contribute to the gravitational index. 
Several supersymmetric D3-brane embeddings are possible in the complexified background, 
 whose action $I$ is still given by \eqref{Ionshell_AdS5}.
In particular, one can place an instantonic D3-brane at the black hole horizon $r=r_+$, wrapping a suitable $S^1$ 
inside the $S^3$ in the asymptotically AdS$_5$ space, as well as
an $S^3$ inside $S^5$.  
Then the D3-brane action evaluates to
\be\label{eq:actionD3}
I_{\rm D3} \= \pm\, \pi N \, \frac{4\varphi}{3\omega_i}\,,
\ee
where 
the upper/lower sign choice refers to D3 or anti-D3 branes and is correlated to the upper/lower sign choice in~\eqref{constraint_chempot_5d}, while the choice $i=1,2$ corresponds to a D3 wrapping either one or the other possible circles in the horizon three-sphere.

Using the brane stability criterion~\eqref{stability_criterion}, one can see that many candidate saddles are in fact disallowed. For instance, consider the family of supersymmetric black hole solutions with angular velocities of the form $\omega_i = \omega + 2\pi \ii n_i$, where $\omega$ is a reference value and $n_i$ are integers. Although they all satisfy the same boundary conditions, using \eqref{eq:actionD3} 
 one can see that the only configurations passing the criterion are those with $n_1=n_2$, and hence with $\omega_1=\omega_2$.
This nicely matches the findings on the SCFT side, where the known Bethe roots of the $\mathcal{N}=4$ SYM index are precisely of the same form.
 Moreover, for the configurations passing the criterion, one also finds that the form of the non-perturbative corrections \eqref{eq:I_corrected} to the leading $\mathcal{O}(N^2)$ term given by the action $I$ does match the one obtained analyzing the subleading contributions to the SCFT index (this refers to the  exponent, while matching the coefficient of the exponential would require further work).

\vskip 4pt

Another discrete transformation discussed in \cite{Aharony:2021zkr} is a specific orbifold of the supersymmetric non-extremal black hole solution. Here the quotient is taken by a $\mathbb{Z}_m$ discrete symmetry embedded in the U(1)$^6$ symmetry of the solution uplifted to type IIB supergravity (the U(1)$^6$ is made of the translations along the closed orbits of the Killing vector generating the horizon, 
of the U(1)$^2$ rotations in the asymptotically AdS$_5$ space, and of the U(1)$^3$ rotations in $S^5$). In particular, the orbifold reduces the period of the Euclidean time circle by a factor of  $1/m$. One can arrange for the orbifold not to break supersymmetry, as well as to satisfy the same boundary conditions as unorbifolded black hole solutions. For simplicity let us consider the case of equal angular velocities, $\omega_1=\omega_2\equiv\omega$. One can see that 
 the $\mathbb{Z}_m$ orbifold of a supersymmetric solution with chemical potentials $(\tilde\beta,\,\tilde\omega,\,\tilde\varphi=\tilde\omega \mp \pi \ii)$ satisfies the same boundary conditions as an unorbifolded solution with chemical potentials $(\beta,\,\omega,\,\varphi)$, provided $(\tilde\beta = m\beta, \tilde\omega = m\omega+2\pi \ii r  , \tilde\varphi = m\varphi +3\pi \ii s)$, for suitable integers $r,s$ satisfying $2r-3s=m\pm 1$ (mod $2m$). Hence the two solutions may contribute to the same gravitational index. The on-shell action of this orbifolded solution has the form~\eqref{Ionshell_AdS5} (in terms of the tilded potentials), but divided by $m$ in order to take into account the reduced Euclidean time period.
It can then be written as
\be \label{eq:orbifoldaction}
I_{\rm orbifold} \= \frac{2\pi}{27G g^3}\,\frac{(m\varphi +3\pi\ii s)^3}{m(m\omega+2\pi \ii r)^2}\,.
\ee
Again one can apply the brane stability criterion to select the admissible saddles to be included in the partition function, and again one finds that both the classical action and the D3-brane non-perturbative corrections match the contribution of some other Bethe roots to the $\mathcal{N}=4$ SYM index.
The same contributions can be found using the elliptic continuation method~\cite{Cabo-Bizet:2019eaf,Cabo-Bizet:2020nkr}, 
as well as by carefully analyzing the novel Cardy-like regimes  obtained when $\omega/2\pi \ii$ approaches the rational points~$-r/m$~\cite{ArabiArdehali:2021nsx}. 

\vskip 4pt

In this way a precise one-to-one correspondence is reached between supersymmetric gravitational solutions passing the stability condition and known 
saddle-points of the dual SCFT index.
We refer to \cite{Aharony:2021zkr,Aharony:2024ntg} for a more thorough comparison of gravitational and SCFT results, also including the chemical potentials for 
the other two independent U(1) isometries of $S^5$, that we did not have the chance to discuss here.

\vskip 4pt

The formula~\eqref{eq:orbifoldaction} shows that there are an infinite number of saddles labelled by integers~$(m,r,s)$ with different actions. 
The BH with~$m=1$ gives the leading growth of microcanonical index (in absolute value), and the~$m=2,3,\dots$ give exponentially subleading contributions. 
On the other hand, at a given point in the space of chemical potentials, 
the saddle with the largest value of the action  dominates the grand-canonical index~\cite{Cabo-Bizet:2019eaf}. 
In particular, when~$\omega/2\pi \ii$ is close enough to the rational point~$-r/m$, the corresponding orbifold 
dominates the ensemble---in spite of the suppression of the action by~$1/m$.
This leads to a rich (and very erratic) phase diagram in the space of chemical potentials~\cite{Cabo-Bizet:2019eaf}.

\section{Supersymmetric black holes in asymptotically flat space} \label{sec:BHsflatspace}

In the previous section we have seen how the general ideas of Section~\ref{sec:grandcanindex} about the gravitational supersymmetric index and its black hole saddles are made concrete in AdS supergravity, and how a precise matching is found in the dual SCFTs. Here we briefly discuss how the same ideas are also useful in the case of asymptotically flat boundary conditions,
although we do not have a dual SCFT to compare with. 
The gravitational saddles for the index find their simplest realization in this set-up, as we see below.

AdS black holes reduce to asymptotically flat black holes by taking the limit where the R-symmetry gauging parameter $g$, which is also the inverse AdS radius, goes to zero. In this ungauged limit 
the supercharges do not transform under any of the electric charges in the theory, i.e.\ the $r_I$ appearing in \eqref{comm_chargesupercharge} all vanish.
It follows that several expressions simplify. For instance, the electrostatic potentials drop out of the supersymmetric boundary condition~\eqref{eq:susyconstraint}, which then reads
\be
\sum_{i}\,\omega^i  \= 2\pi  \ii n\,,\qquad n\in\mathbb{Z}_{\rm odd}\,.
\ee
Solving for one of the angular velocities, say $\omega^1$, and using  $\rme^{2\pi \ii J_1} = (-1)^\FF$, allows us  to express 
 the partition function~\eqref{eq:Z_almostsusy} as the index
\be
Z(\omega^i,\varphi^I) 
\= {\rm Tr} \,(-1)^\FF \rme^{-\beta \{\mathcal{Q},\mathcal{Q}^\dagger\}  +\sum_{i\neq 1}\omega^i (J_i-J_1) +\varphi^I Q_I}\,.
\ee
A related simplification is that the metric of the asymptotically flat solution admits a real slice, which simplifies the regularity analysis, however it may also be taken complex. 
Also, at least one combination of the  angular momenta is imaginary.

In particular, in five dimensions the index depends on one angular velocity and the $\varphi^I$, and counts states with charges $J_2-J_1$ and $Q_I$. 
The general rotating and charged black hole solution of minimal gauged supergravity discussed in Section~\ref{sec:aAdS_blackholes} reduces in the ungauged limit\footnote{We specified in footnote~\ref{foot:conventions_gauge} how this limit must be taken in the expressions of Section~\ref{sec:asympt_AdS}.} to a solution originally found in~\cite{Cvetic:1996xz}, which includes the BMPV black hole~\cite{Breckenridge:1996is}. The solution admits a Killing spinor (while remaining non-extremal in general) when the angular velocities satisfy the condition
\be
\omega_1+\omega_2 \= \pm 2\pi \ii\,,
\ee
and contributes to the gravitational index via a saddle point action which is still given by~\eqref{Ionshell_AdS5}. The Legendre transform of this action now gives an expression for the Bekenstein-Hawking entropy which is real if  $J_1+J_2=0$, and precisely matches the entropy of the supersymmetric and extremal BMPV black hole. This saddle of the index and its generalizations to multiple electric charges and other topologies, as well as quantum corrections, have been recently studied in~\cite{Anupam:2023yns,Hegde:2023jmp,Cassani:2024kjn,Boruch:2025qdq}. 

\vskip 4pt

We can also come back to four-dimensional ungauged supergravity and provide a complementary 
discussion to that in Section~\ref{sec:QuantumEntropyBH}
from the viewpoint of asymptotically flat space. 
Implementing the ideas above in the framework of four-dimensional pure $\mathcal{N}=2$ supergravity in fact provides
the simplest example of a gravitational supersymmetric index and its finite-temperature saddles~\cite{Iliesiu:2021are,Hristov:2022pmo,H:2023qko,Boruch:2023gfn}, so we find it instructive to illustrate it in some detail. 

The bosonic action of pure $\mathcal{N}=2$ supergravity was given in~\eqref{eq:EMaction} and we now consider it in Euclidean signature. 
 The supersymmetric boundary condition requires the only angular velocity to satisfy
 \be
 \omega \;\equiv\; \beta \Omega\=  2\pi \ii n \,,\qquad n\in\mathbb{Z}_{\rm odd}\,,
 \ee
 so that $\rme^{\omega J} = (-1)^\FF $,
 and the grand-canonical partition function takes the form
\be
Z  \= {\rm Tr}\, (-1)^\FF  \rme^{-\beta \{\mathcal{Q}, {\mathcal{Q}^\dagger} \} + \varphi Q}\,.
\ee
 Let us first consider the Euclidean Kerr-Newman black hole solution. The finite $\beta$ solution with
 $\omega = \pm 2\pi \ii$ is supersymmetric
 and contributes to the gravitational index (in terms of the conserved charges, this supersymmetric solution has $M=Q$ and only becomes extremal if $J=0$). However, in this case we can go further and study more general saddles. 
The Lorentzian supersymmetric solutions were classified long ago in~\cite{Tod:1983pm},
and, in fact, the resulting solutions 
in the so-called timelike class go back even further to the work of Israel-Wilson and Perj\'es~\cite{Israel:1972vx,Perjes:1971gv}. 
Their Euclidean counterpart  were studied in \cite{Whitt:1984wk,Yuille:1987vw} and the recent insight has led to interpret them as the semiclassical saddles of the gravitational index.
 The metric and gauge field read
\begin{equation}
\begin{aligned}
\label{eq:IWPmetric}
\diff s^2& \= \big( V\widetilde{ V}\big)^{-1}\left( \diff \tau + \breve\omega\right)^2 +  V\widetilde{ V}\, \diff {\bf x} \cdot \diff {\bf x}\,,\\[1mm]
 A &\= \frac{\ii}{2}\,\big(V^{-1}+\widetilde{V}^{-1}\big)\left( \diff \tau + \breve\omega\right) + \breve A \,, 
\end{aligned}
\end{equation}
where $V,\widetilde{V}$ are independent, real harmonic functions on a three-dimensional flat base space parameterized by ${\bf x}=(x^1,x^2,x^3)$, and
  $\breve\omega$, $\breve A$ are local one-forms on the base given by
\begin{equation}
\begin{aligned}
*_3\, \diff \breve\omega \= \big(\,\widetilde{V}\,\diff  V- V\, \diff \widetilde{ V}\,\big) \,\,,\qquad\qquad 
*_3\,\diff\breve A \= -\frac{\ii}{2}\,\diff\,\big(\, V - \widetilde{V}\,\big)  \,.
\end{aligned}
\end{equation}

In particular, the Euclidean supersymmetric non-extremal Kerr-Newman solution in these coordinates is realized  by choosing the harmonic functions as
\begin{equation}\label{eq:choiceVtildeV}
 V \= 1 + \frac{Q}{|{\bf x}-{\bf x}_N|}\,\,,\qquad\qquad \widetilde{ V} \= 1 + \frac{Q}{|{\bf x}-{\bf x}_S|}\,,
\end{equation}
where ${\bf x}_N$ and ${\bf x}_S$ are two points in the base, denoted as centers, and $Q$ is the electric charge.  The inverse temperature $\beta$ and the redefined electrostatic potential $\varphi$ read
\begin{equation}\label{eq:beta_Phi_4D}
\beta \= \frac{4\pi Q\left( Q+\delta\right)}{\delta}\,,\qquad\qquad
\varphi \;\equiv\; \beta\big( \Phi - 1\,\big) \= -2\pi Q \,.
\end{equation}
where $\delta = |{\bf x}_N - {\bf x}_S|$ is the distance between the centers. The extremal solution is obtained by taking the two centers coincide, $\delta\to 0$, which implies $V=\widetilde V$ and therefore $\breve{\omega}=\breve{A}=0$, which turns off the rotation. The axis between the two centers is mapped into the horizon two-sphere in Boyer-Lindquist coordinates.
The supersymmetric on-shell action with Dirichlet boundary conditions for both the metric and the gauge field reads
\begin{equation}
I\= \frac{\varphi^2}{4\pi}\,,
\end{equation}
and gives the saddle-point contribution of the supersymmetric Kerr-Newman solution to the grand-canonical index.

More general solutions can easily be constructed by choosing harmonic functions with multiple centers, which leads to solutions with additional two-cycles. 
A  study of the global properties of these solutions has the potential to provide a complete characterization of the saddles of the gravitational index. 

 Also, the above ideas presented for pure~$\CN=2$ supergravity extend in a nice manner to 
more general theories of four-dimensional $\CN=2$ supergravity coupled to an arbitrary number of vector multiplets~\cite{Boruch:2023gfn}. 
The rotating supersymmetric BH solutions are the true saddles of the gravitational  index. 
Unlike the attractor BHs~\cite{Ferrara:1995ih}, they are not spherically symmetric and they carry 
electromagnetic dipole fields strengths for the vector fields.
The scalar fields are not constant on the horizon and depend on the moduli values at infinity and the temperature. 
The horizon area is also a function of the moduli and the temperature. 
Nevertheless, the scalar fields at the north and south poles of the horizon two-sphere are fixed completely in terms 
of the charges, and the free energy of the BH agrees precisely with that of the extremal BH. 
These properties are summarized in terms of the \emph{new attractor mechanism} explained in~\cite{Boruch:2023gfn}.

\section{Future directions}

The developments described in this article lead to many interesting directions of research, many of which we did not discuss.

The classification of saddle-points that contribute to the gravitational path integral 
is an important issue. 
Explicit saddle-points allow us to concretely flesh out the ideas of the gravitational index,
as well as serving as important starting points for the calculation of quantum effects. 
However, obtaining all explicit solutions in general problems can be challenging. 
The methods of equivariant localization as applied to calculate classical supergravity actions and other quantities~\cite{BenettiGenolini:2023kxp,Martelli:2023oqk,BenettiGenolini:2023ndb,BenettiGenolini:2024kyy}   
may allow us to circumvent this problem in an elegant manner, in the spirit of the nuts and bolts of Gibbons-Hawking~\cite{Gibbons:1979xm}.

The idea of the gravitational index and the techniques of supersymmetric localization
can be applied to include defects and branes. 
Along these lines, the action of various wrapped D3-branes in the complexified supersymmetric black hole background was studied in~\cite{Aharony:2021zkr} for instantonic D-branes and~\cite{Chen:2023lzq} for D3-brane configurations realizing Gukov-Witten defects. 
In all these cases the D-brane action takes a nice form in terms of the chemical potentials.

The appearance of complex saddles of the gravitational path integral 
naturally leads to the question of which saddles are allowable
and, relatedly, what is the correct contour of integration? 
Interesting ideas have arisen from different viewpoints such as the role of multi-gravitons~\cite{Cabo-Bizet:2020ewf}, 
D-brane stability~\cite{Aharony:2021zkr}  
and the Kontsevich-Segal-Witten criterion~\cite{Kontsevich:2021dmb,Witten:2021nzp}.

Some of the calculations of quantum corrections to BH entropy can be extended to near- and non-extremal situations, e.g.~the leading logarithmic effects 
in the horizon area~\cite{Sen:2012dw} 
and in the temperature~\cite{Iliesiu:2020qvm}. 
Another type of correction comes from higher-derivative terms, for which it would be interesting to nail down the higher-dimensional origin and to prove non-renormalization theorems in supergravity. 
All these corrections provide concrete 
low-energy constraints for any UV theory of quantum gravity, as well as being potentially important for applied holography.

It would be nice to extend the ideas of exact quantum BH entropy to diverse dimensions~\cite{Gupta:2021roy},
to~$\CN=4$ supergravity where we also have exact number-theoretic results~\cite{Dabholkar:2012nd, Ferrari:2017msn, Bhand:2023rhm},
and, more generally, to~$\CN=2$ supergravity~\cite{Denef:2007vg}. 
At a more formal level, the calculation of the localized supersymmetric gravitational path integral 
can be thought of as a starting point of a rigorous theory of twisted quantum supergravity~\cite{Costello:2016mgj, deWit:2018dix}.
One important direction along these lines is to consider gauged supergravity where we need to include hypermultiplets (see e.g.~\cite{Dabholkar:2014wpa,Hristov:2018lod,Hristov:2021zai}).  
Doing so would allow us to define the notion of exact holography where we can calculate 
both sides of the holographic correspondence with analytic control.

\section*{Acknowledgments}

We would like to thank the editors, Anna Ceresole and Gianguido Dall'Agata, for inviting us to prepare the present contribution. We also thank 
Jan Boruch, 
Pietro Benetti Genolini, 
Alejandro Cabo-Bizet, 
Atish Dabholkar, 
Bernard de Wit, 
Roberto Emparan, 
Jo\~ao Gomes, 
Rajesh Gupta, 
Luca Iliesiu,  
Imtak Jeon, 
Dario Martelli, 
Valentin Reys, 
Alejandro Ruip\'erez, 
Manya Sahni, 
Enrico Turetta, and 
Joaquin Turiaci
for collaboration on the topics discussed here.

\setstretch{0.9}

\bibliography{Sugra50QBHLocalization.bib}

\providecommand{\href}[2]{#2}\begingroup\raggedright\begin{thebibliography}{100}

\bibitem{Bekenstein:1973ur}
J.D.~Bekenstein, \emph{{Black holes and entropy}},
  \href{https://doi.org/10.1103/PhysRevD.7.2333}{\emph{Phys. Rev. D} {\bfseries
  7} (1973) 2333}.

\bibitem{Hawking:1975vcx}
S.W.~Hawking, \emph{{Particle Creation by Black Holes}},
  \href{https://doi.org/10.1007/BF02345020}{\emph{Commun. Math. Phys.}
  {\bfseries 43} (1975) 199}.

\bibitem{Strominger:1996sh}
A.~Strominger and C.~Vafa, \emph{{Microscopic origin of the Bekenstein-Hawking
  entropy}}, \href{https://doi.org/10.1016/0370-2693(96)00345-0}{\emph{Phys.
  Lett.} {\bfseries B379} (1996) 99}
  [\href{https://arxiv.org/abs/hep-th/9601029}{{\ttfamily hep-th/9601029}}].

\bibitem{Mohaupt:2000mj}
T.~Mohaupt, \emph{{Black hole entropy, special geometry and strings}},
  \href{https://doi.org/10.1002/1521-3978(200102)49:1/3<3::AID-PROP3>3.0.CO;2-#}{\emph{Fortsch.
  Phys.} {\bfseries 49} (2001) 3}
  [\href{https://arxiv.org/abs/hep-th/0007195}{{\ttfamily hep-th/0007195}}].

\bibitem{Sen:2007qy}
A.~Sen, \emph{{Black Hole Entropy Function, Attractors and Precision Counting
  of Microstates}}, \href{https://doi.org/10.1007/s10714-008-0626-4}{\emph{Gen.
  Rel. Grav.} {\bfseries 40} (2008) 2249}
  [\href{https://arxiv.org/abs/0708.1270}{{\ttfamily 0708.1270}}].

\bibitem{Sen:2008vm}
A.~Sen, \emph{{Quantum Entropy Function from AdS(2)/CFT(1) Correspondence}},
  \href{https://doi.org/10.1142/S0217751X09045893}{\emph{Int. J. Mod. Phys. A}
  {\bfseries 24} (2009) 4225}
  [\href{https://arxiv.org/abs/0809.3304}{{\ttfamily 0809.3304}}].

\bibitem{Iliesiu:2020qvm}
L.V.~Iliesiu and G.J.~Turiaci, \emph{{The statistical mechanics of
  near-extremal black holes}},
  \href{https://doi.org/10.1007/JHEP05(2021)145}{\emph{JHEP} {\bfseries 05}
  (2021) 145} [\href{https://arxiv.org/abs/2003.02860}{{\ttfamily
  2003.02860}}].

\bibitem{Iliesiu:2022kny}
L.V.~Iliesiu, S.~Murthy and G.J.~Turiaci, \emph{{Black hole microstate counting
  from the gravitational path integral}},
  \href{https://arxiv.org/abs/2209.13602}{{\ttfamily 2209.13602}}.

\bibitem{Gibbons:1976ue}
G.W.~Gibbons and S.W.~Hawking, \emph{{Action Integrals and Partition Functions
  in Quantum Gravity}},
  \href{https://doi.org/10.1103/PhysRevD.15.2752}{\emph{Phys. Rev.} {\bfseries
  D15} (1977) 2752}.

\bibitem{Gibbons:1978ac}
G.W.~Gibbons, S.W.~Hawking and M.J.~Perry, \emph{{Path Integrals and the
  Indefiniteness of the Gravitational Action}},
  \href{https://doi.org/10.1016/0550-3213(78)90161-X}{\emph{Nucl. Phys. B}
  {\bfseries 138} (1978) 141}.

\bibitem{Hawking:1982dh}
S.W.~Hawking and D.N.~Page, \emph{{Thermodynamics of Black Holes in anti-De
  Sitter Space}}, \href{https://doi.org/10.1007/BF01208266}{\emph{Commun. Math.
  Phys.} {\bfseries 87} (1983) 577}.

\bibitem{Witten:1998zw}
E.~Witten, \emph{{Anti-de Sitter space, thermal phase transition, and
  confinement in gauge theories}},
  \href{https://doi.org/10.4310/ATMP.1998.v2.n3.a3}{\emph{Adv. Theor. Math.
  Phys.} {\bfseries 2} (1998) 505}
  [\href{https://arxiv.org/abs/hep-th/9803131}{{\ttfamily hep-th/9803131}}].

\bibitem{Aharony:1999ti}
O.~Aharony, S.S.~Gubser, J.M.~Maldacena, H.~Ooguri and Y.~Oz, \emph{{Large N
  field theories, string theory and gravity}},
  \href{https://doi.org/10.1016/S0370-1573(99)00083-6}{\emph{Phys. Rept.}
  {\bfseries 323} (2000) 183}
  [\href{https://arxiv.org/abs/hep-th/9905111}{{\ttfamily hep-th/9905111}}].

\bibitem{Witten:1982df}
E.~Witten, \emph{{Constraints on Supersymmetry Breaking}},
  \href{https://doi.org/10.1016/0550-3213(82)90071-2}{\emph{Nucl. Phys. B}
  {\bfseries 202} (1982) 253}.

\bibitem{Wald:1993nt}
R.M.~Wald, \emph{{Black hole entropy is the Noether charge}},
  \href{https://doi.org/10.1103/PhysRevD.48.R3427}{\emph{Phys. Rev. D}
  {\bfseries 48} (1993) R3427}
  [\href{https://arxiv.org/abs/gr-qc/9307038}{{\ttfamily gr-qc/9307038}}].

\bibitem{Iyer:1994ys}
V.~Iyer and R.M.~Wald, \emph{{Some properties of Noether charge and a proposal
  for dynamical black hole entropy}},
  \href{https://doi.org/10.1103/PhysRevD.50.846}{\emph{Phys. Rev. D} {\bfseries
  50} (1994) 846} [\href{https://arxiv.org/abs/gr-qc/9403028}{{\ttfamily
  gr-qc/9403028}}].

\bibitem{Freedman:2012zz}
D.Z.~Freedman and A.~Van~Proeyen, \emph{{Supergravity}}, Cambridge Univ. Press,
  Cambridge, UK (5, 2012),
  \href{https://doi.org/10.1017/CBO9781139026833}{10.1017/CBO9781139026833}.

\bibitem{deWit:1980lyi}
B.~de~Wit, J.W.~van Holten and A.~Van~Proeyen, \emph{{Structure of N=2
  Supergravity}},
  \href{https://doi.org/10.1016/0550-3213(83)90548-5}{\emph{Nucl. Phys. B}
  {\bfseries 184} (1981) 77}.

\bibitem{Bodner:1990zm}
M.~Bodner, A.C.~Cadavid and S.~Ferrara, \emph{{(2,2) vacuum configurations for
  type IIA superstrings: N=2 supergravity Lagrangians and algebraic geometry}},
  \href{https://doi.org/10.1088/0264-9381/8/5/005}{\emph{Class. Quant. Grav.}
  {\bfseries 8} (1991) 789}.

\bibitem{Ferrara:1995ih}
S.~Ferrara, R.~Kallosh and A.~Strominger, \emph{{N=2 extremal black holes}},
  \href{https://doi.org/10.1103/PhysRevD.52.R5412}{\emph{Phys. Rev. D}
  {\bfseries 52} (1995) R5412}
  [\href{https://arxiv.org/abs/hep-th/9508072}{{\ttfamily hep-th/9508072}}].

\bibitem{Sen:2012kpz}
A.~Sen, \emph{{Logarithmic Corrections to N=2 Black Hole Entropy: An Infrared
  Window into the Microstates}},
  \href{https://doi.org/10.1007/s10714-012-1336-5}{\emph{Gen. Rel. Grav.}
  {\bfseries 44} (2012) 1207}
  [\href{https://arxiv.org/abs/1108.3842}{{\ttfamily 1108.3842}}].

\bibitem{LopesCardoso:1998tkj}
G.~Lopes~Cardoso, B.~de~Wit and T.~Mohaupt, \emph{{Corrections to macroscopic
  supersymmetric black hole entropy}},
  \href{https://doi.org/10.1016/S0370-2693(99)00227-0}{\emph{Phys. Lett. B}
  {\bfseries 451} (1999) 309}
  [\href{https://arxiv.org/abs/hep-th/9812082}{{\ttfamily hep-th/9812082}}].

\bibitem{LopesCardoso:2000qm}
G.~Lopes~Cardoso, B.~de~Wit, J.~Kappeli and T.~Mohaupt, \emph{{Stationary BPS
  solutions in N=2 supergravity with R**2 interactions}},
  \href{https://doi.org/10.1088/1126-6708/2000/12/019}{\emph{JHEP} {\bfseries
  12} (2000) 019} [\href{https://arxiv.org/abs/hep-th/0009234}{{\ttfamily
  hep-th/0009234}}].

\bibitem{Ooguri:2004zv}
H.~Ooguri, A.~Strominger and C.~Vafa, \emph{{Black hole attractors and the
  topological string}},
  \href{https://doi.org/10.1103/PhysRevD.70.106007}{\emph{Phys. Rev. D}
  {\bfseries 70} (2004) 106007}
  [\href{https://arxiv.org/abs/hep-th/0405146}{{\ttfamily hep-th/0405146}}].

\bibitem{Banerjee:2010qc}
S.~Banerjee, R.K.~Gupta and A.~Sen, \emph{{Logarithmic Corrections to Extremal
  Black Hole Entropy from Quantum Entropy Function}},
  \href{https://doi.org/10.1007/JHEP03(2011)147}{\emph{JHEP} {\bfseries 03}
  (2011) 147} [\href{https://arxiv.org/abs/1005.3044}{{\ttfamily 1005.3044}}].

\bibitem{Banerjee:2011jp}
S.~Banerjee, R.K.~Gupta, I.~Mandal and A.~Sen, \emph{{Logarithmic Corrections
  to N=4 and N=8 Black Hole Entropy: A One Loop Test of Quantum Gravity}},
  \href{https://doi.org/10.1007/JHEP11(2011)143}{\emph{JHEP} {\bfseries 11}
  (2011) 143} [\href{https://arxiv.org/abs/1106.0080}{{\ttfamily 1106.0080}}].

\bibitem{Dabholkar:2010uh}
A.~Dabholkar, J.~Gomes and S.~Murthy, \emph{{Quantum black holes, localization
  and the topological string}},
  \href{https://doi.org/10.1007/JHEP06(2011)019}{\emph{JHEP} {\bfseries 06}
  (2011) 019} [\href{https://arxiv.org/abs/1012.0265}{{\ttfamily 1012.0265}}].

\bibitem{Dabholkar:2011ec}
A.~Dabholkar, J.~Gomes and S.~Murthy, \emph{{Localization \& Exact
  Holography}}, \href{https://doi.org/10.1007/JHEP04(2013)062}{\emph{JHEP}
  {\bfseries 04} (2013) 062} [\href{https://arxiv.org/abs/1111.1161}{{\ttfamily
  1111.1161}}].

\bibitem{Witten:1988ze}
E.~Witten, \emph{{Topological Quantum Field Theory}},
  \href{https://doi.org/10.1007/BF01223371}{\emph{Commun. Math. Phys.}
  {\bfseries 117} (1988) 353}.

\bibitem{Nekrasov:2002qd}
N.A.~Nekrasov, \emph{{Seiberg-Witten prepotential from instanton counting}},
  \href{https://doi.org/10.4310/ATMP.2003.v7.n5.a4}{\emph{Adv. Theor. Math.
  Phys.} {\bfseries 7} (2003) 831}
  [\href{https://arxiv.org/abs/hep-th/0206161}{{\ttfamily hep-th/0206161}}].

\bibitem{Pestun:2007rz}
V.~Pestun, \emph{{Localization of gauge theory on a four-sphere and
  supersymmetric Wilson loops}},
  \href{https://doi.org/10.1007/s00220-012-1485-0}{\emph{Commun. Math. Phys.}
  {\bfseries 313} (2012) 71} [\href{https://arxiv.org/abs/0712.2824}{{\ttfamily
  0712.2824}}].

\bibitem{Pestun:2016zxk}
V.~Pestun et~al., \emph{{Localization techniques in quantum field theories}},
  \href{https://doi.org/10.1088/1751-8121/aa63c1}{\emph{J. Phys. A} {\bfseries
  50} (2017) 440301} [\href{https://arxiv.org/abs/1608.02952}{{\ttfamily
  1608.02952}}].

\bibitem{deWit:2018dix}
B.~de~Wit, S.~Murthy and V.~Reys, \emph{{BRST quantization and equivariant
  cohomology: localization with asymptotic boundaries}},
  \href{https://doi.org/10.1007/JHEP09(2018)084}{\emph{JHEP} {\bfseries 09}
  (2018) 084} [\href{https://arxiv.org/abs/1806.03690}{{\ttfamily
  1806.03690}}].

\bibitem{Jeon:2018kec}
I.~Jeon and S.~Murthy, \emph{{Twisting and localization in supergravity:
  equivariant cohomology of BPS black holes}},
  \href{https://doi.org/10.1007/JHEP03(2019)140}{\emph{JHEP} {\bfseries 03}
  (2019) 140} [\href{https://arxiv.org/abs/1806.04479}{{\ttfamily
  1806.04479}}].

\bibitem{Baulieu:1988xs}
L.~Baulieu and I.M.~Singer, \emph{{Topological Yang-Mills symmetry}},
  \href{https://doi.org/10.1016/0920-5632(88)90366-0}{\emph{Nucl. Phys. B Proc.
  Suppl.} {\bfseries 5} (1988) 12}.

\bibitem{Costello:2016mgj}
K.~Costello and S.~Li, \emph{{Twisted supergravity and its quantization}},
  \href{https://arxiv.org/abs/1606.00365}{{\ttfamily 1606.00365}}.

\bibitem{Gupta:2012cy}
R.K.~Gupta and S.~Murthy, \emph{{All solutions of the localization equations
  for N=2 quantum black hole entropy}},
  \href{https://doi.org/10.1007/JHEP02(2013)141}{\emph{JHEP} {\bfseries 02}
  (2013) 141} [\href{https://arxiv.org/abs/1208.6221}{{\ttfamily 1208.6221}}].

\bibitem{Tod:1983pm}
K.P.~Tod, \emph{{All Metrics Admitting Supercovariantly Constant Spinors}},
  \href{https://doi.org/10.1016/0370-2693(83)90797-9}{\emph{Phys. Lett. B}
  {\bfseries 121} (1983) 241}.

\bibitem{Gauntlett:2002nw}
J.P.~Gauntlett, J.B.~Gutowski, C.M.~Hull, S.~Pakis and H.S.~Reall, \emph{{All
  supersymmetric solutions of minimal supergravity in five- dimensions}},
  \href{https://doi.org/10.1088/0264-9381/20/21/005}{\emph{Class. Quant. Grav.}
  {\bfseries 20} (2003) 4587}
  [\href{https://arxiv.org/abs/hep-th/0209114}{{\ttfamily hep-th/0209114}}].

\bibitem{Antoniadis:1993ze}
I.~Antoniadis, E.~Gava, K.S.~Narain and T.R.~Taylor, \emph{{Topological
  amplitudes in string theory}},
  \href{https://doi.org/10.1016/0550-3213(94)90617-3}{\emph{Nucl. Phys. B}
  {\bfseries 413} (1994) 162}
  [\href{https://arxiv.org/abs/hep-th/9307158}{{\ttfamily hep-th/9307158}}].

\bibitem{Bershadsky:1993cx}
M.~Bershadsky, S.~Cecotti, H.~Ooguri and C.~Vafa, \emph{{Kodaira-Spencer theory
  of gravity and exact results for quantum string amplitudes}},
  \href{https://doi.org/10.1007/BF02099774}{\emph{Commun. Math. Phys.}
  {\bfseries 165} (1994) 311}
  [\href{https://arxiv.org/abs/hep-th/9309140}{{\ttfamily hep-th/9309140}}].

\bibitem{deWit:2010za}
B.~de~Wit, S.~Katmadas and M.~van Zalk, \emph{{New supersymmetric
  higher-derivative couplings: Full N=2 superspace does not count!}},
  \href{https://doi.org/10.1007/JHEP01(2011)007}{\emph{JHEP} {\bfseries 01}
  (2011) 007} [\href{https://arxiv.org/abs/1010.2150}{{\ttfamily 1010.2150}}].

\bibitem{Butter:2013lta}
D.~Butter, B.~de~Wit, S.M.~Kuzenko and I.~Lodato, \emph{{New higher-derivative
  invariants in N=2 supergravity and the Gauss-Bonnet term}},
  \href{https://doi.org/10.1007/JHEP12(2013)062}{\emph{JHEP} {\bfseries 12}
  (2013) 062} [\href{https://arxiv.org/abs/1307.6546}{{\ttfamily 1307.6546}}].

\bibitem{Murthy:2013xpa}
S.~Murthy and V.~Reys, \emph{{Quantum black hole entropy and the holomorphic
  prepotential of N=2 supergravity}},
  \href{https://doi.org/10.1007/JHEP10(2013)099}{\emph{JHEP} {\bfseries 10}
  (2013) 099} [\href{https://arxiv.org/abs/1306.3796}{{\ttfamily 1306.3796}}].

\bibitem{Butter:2014iwa}
D.~Butter, B.~de~Wit and I.~Lodato, \emph{{Non-renormalization theorems and N=2
  supersymmetric backgrounds}},
  \href{https://doi.org/10.1007/JHEP03(2014)131}{\emph{JHEP} {\bfseries 03}
  (2014) 131} [\href{https://arxiv.org/abs/1401.6591}{{\ttfamily 1401.6591}}].

\bibitem{David:2018pex}
J.R.~David, E.~Gava, R.K.~Gupta and K.~Narain, \emph{{Boundary conditions and
  localization on AdS. Part I}},
  \href{https://doi.org/10.1007/JHEP09(2018)063}{\emph{JHEP} {\bfseries 09}
  (2018) 063} [\href{https://arxiv.org/abs/1802.00427}{{\ttfamily
  1802.00427}}].

\bibitem{David:2019ocd}
J.R.~David, E.~Gava, R.K.~Gupta and K.~Narain, \emph{{Boundary conditions and
  localization on AdS. Part II. General analysis}},
  \href{https://doi.org/10.1007/JHEP02(2020)139}{\emph{JHEP} {\bfseries 02}
  (2020) 139} [\href{https://arxiv.org/abs/1906.02722}{{\ttfamily
  1906.02722}}].

\bibitem{Murthy:2015yfa}
S.~Murthy and V.~Reys, \emph{{Functional determinants, index theorems, and
  exact quantum black hole entropy}},
  \href{https://doi.org/10.1007/JHEP12(2015)028}{\emph{JHEP} {\bfseries 12}
  (2015) 028} [\href{https://arxiv.org/abs/1504.01400}{{\ttfamily
  1504.01400}}].

\bibitem{Sen:2023dps}
A.~Sen, \emph{{Revisiting localization for BPS black hole entropy}},
  \href{https://arxiv.org/abs/2302.13490}{{\ttfamily 2302.13490}}.

\bibitem{GonzalezLezcano:2023uar}
A.~Gonz\'alez~Lezcano, I.~Jeon and A.~Ray, \emph{{Supersymmetry and
  complexified spectrum on Euclidean AdS2}},
  \href{https://doi.org/10.1103/PhysRevD.108.045018}{\emph{Phys. Rev. D}
  {\bfseries 108} (2023) 045018}
  [\href{https://arxiv.org/abs/2305.12925}{{\ttfamily 2305.12925}}].

\bibitem{Hristov:2019xku}
K.~Hristov, I.~Lodato and V.~Reys, \emph{{One-loop determinants for black holes
  in 4d gauged supergravity}},
  \href{https://doi.org/10.1007/JHEP11(2019)105}{\emph{JHEP} {\bfseries 11}
  (2019) 105} [\href{https://arxiv.org/abs/1908.05696}{{\ttfamily
  1908.05696}}].

\bibitem{Dabholkar:2014ema}
A.~Dabholkar, J.~Gomes and S.~Murthy, \emph{{Nonperturbative black hole entropy
  and Kloosterman sums}},
  \href{https://doi.org/10.1007/JHEP03(2015)074}{\emph{JHEP} {\bfseries 03}
  (2015) 074} [\href{https://arxiv.org/abs/1404.0033}{{\ttfamily 1404.0033}}].

\bibitem{Sen:2009vz}
A.~Sen, \emph{{Arithmetic of Quantum Entropy Function}},
  \href{https://doi.org/10.1088/1126-6708/2009/08/068}{\emph{JHEP} {\bfseries
  08} (2009) 068} [\href{https://arxiv.org/abs/0903.1477}{{\ttfamily
  0903.1477}}].

\bibitem{Dabholkar:2010rm}
A.~Dabholkar, J.~Gomes, S.~Murthy and A.~Sen, \emph{{Supersymmetric Index from
  Black Hole Entropy}},
  \href{https://doi.org/10.1007/JHEP04(2011)034}{\emph{JHEP} {\bfseries 04}
  (2011) 034} [\href{https://arxiv.org/abs/1009.3226}{{\ttfamily 1009.3226}}].

\bibitem{Camporesi:1995fb}
R.~Camporesi and A.~Higuchi, \emph{{On the Eigen functions of the Dirac
  operator on spheres and real hyperbolic spaces}},
  \href{https://doi.org/10.1016/0393-0440(95)00042-9}{\emph{J. Geom. Phys.}
  {\bfseries 20} (1996) 1}
  [\href{https://arxiv.org/abs/gr-qc/9505009}{{\ttfamily gr-qc/9505009}}].

\bibitem{Sachdev:2015efa}
S.~Sachdev, \emph{{Bekenstein-Hawking Entropy and Strange Metals}},
  \href{https://doi.org/10.1103/PhysRevX.5.041025}{\emph{Phys. Rev. X}
  {\bfseries 5} (2015) 041025}
  [\href{https://arxiv.org/abs/1506.05111}{{\ttfamily 1506.05111}}].

\bibitem{Almheiri:2016fws}
A.~Almheiri and B.~Kang, \emph{{Conformal Symmetry Breaking and Thermodynamics
  of Near-Extremal Black Holes}},
  \href{https://doi.org/10.1007/JHEP10(2016)052}{\emph{JHEP} {\bfseries 10}
  (2016) 052} [\href{https://arxiv.org/abs/1606.04108}{{\ttfamily
  1606.04108}}].

\bibitem{Maldacena:2016upp}
J.~Maldacena, D.~Stanford and Z.~Yang, \emph{{Conformal symmetry and its
  breaking in two dimensional Nearly Anti-de-Sitter space}},
  \href{https://doi.org/10.1093/ptep/ptw124}{\emph{PTEP} {\bfseries 2016}
  (2016) 12C104} [\href{https://arxiv.org/abs/1606.01857}{{\ttfamily
  1606.01857}}].

\bibitem{Stanford:2017thb}
D.~Stanford and E.~Witten, \emph{{Fermionic Localization of the Schwarzian
  Theory}}, \href{https://doi.org/10.1007/JHEP10(2017)008}{\emph{JHEP}
  {\bfseries 10} (2017) 008}
  [\href{https://arxiv.org/abs/1703.04612}{{\ttfamily 1703.04612}}].

\bibitem{Moitra:2018jqs}
U.~Moitra, S.P.~Trivedi and V.~Vishal, \emph{{Extremal and near-extremal black
  holes and near-CFT$_{1}$}},
  \href{https://doi.org/10.1007/JHEP07(2019)055}{\emph{JHEP} {\bfseries 07}
  (2019) 055} [\href{https://arxiv.org/abs/1808.08239}{{\ttfamily
  1808.08239}}].

\bibitem{Heydeman:2020hhw}
M.~Heydeman, L.V.~Iliesiu, G.J.~Turiaci and W.~Zhao, \emph{{The statistical
  mechanics of near-BPS black holes}},
  \href{https://doi.org/10.1088/1751-8121/ac3be9}{\emph{J. Phys. A} {\bfseries
  55} (2022) 014004} [\href{https://arxiv.org/abs/2011.01953}{{\ttfamily
  2011.01953}}].

\bibitem{Iliesiu:2022onk}
L.V.~Iliesiu, S.~Murthy and G.J.~Turiaci, \emph{{Revisiting the Logarithmic
  Corrections to the Black Hole Entropy}},
  \href{https://arxiv.org/abs/2209.13608}{{\ttfamily 2209.13608}}.

\bibitem{Preskill:1991tb}
J.~Preskill, P.~Schwarz, A.D.~Shapere, S.~Trivedi and F.~Wilczek,
  \emph{{Limitations on the statistical description of black holes}},
  \href{https://doi.org/10.1142/S0217732391002773}{\emph{Mod. Phys. Lett. A}
  {\bfseries 6} (1991) 2353}.

\bibitem{Murthy:2023mbc}
S.~Murthy, \emph{{Black holes and modular forms in string theory, in Oxford
  Research Encyclopedia of Physics,}}
  \href{https://doi.org/10.1093/acrefore/9780190871994.013.65}{10.1093/acrefore/9780190871994.013.65},
  [\href{https://arxiv.org/abs/2305.11732}{{\ttfamily 2305.11732}}].

\bibitem{Maldacena:1999bp}
J.M.~Maldacena, G.W.~Moore and A.~Strominger, \emph{{Counting BPS black holes
  in toroidal Type II string theory}},
  \href{https://arxiv.org/abs/hep-th/9903163}{{\ttfamily hep-th/9903163}}.

\bibitem{Romans:1991nq}
L.J.~Romans, \emph{{Supersymmetric, cold and lukewarm black holes in
  cosmological Einstein-Maxwell theory}},
  \href{https://doi.org/10.1016/0550-3213(92)90684-4}{\emph{Nucl. Phys. B}
  {\bfseries 383} (1992) 395}
  [\href{https://arxiv.org/abs/hep-th/9203018}{{\ttfamily hep-th/9203018}}].

\bibitem{Cacciatori:2009iz}
S.L.~Cacciatori and D.~Klemm, \emph{{Supersymmetric AdS(4) black holes and
  attractors}}, \href{https://doi.org/10.1007/JHEP01(2010)085}{\emph{JHEP}
  {\bfseries 01} (2010) 085} [\href{https://arxiv.org/abs/0911.4926}{{\ttfamily
  0911.4926}}].

\bibitem{Silva:2006xv}
P.J.~Silva, \emph{{Thermodynamics at the BPS bound for Black Holes in AdS}},
  \href{https://doi.org/10.1088/1126-6708/2006/10/022}{\emph{JHEP} {\bfseries
  10} (2006) 022} [\href{https://arxiv.org/abs/hep-th/0607056}{{\ttfamily
  hep-th/0607056}}].

\bibitem{Iliesiu:2021are}
L.V.~Iliesiu, M.~Kologlu and G.J.~Turiaci, \emph{{Supersymmetric indices
  factorize}}, \href{https://doi.org/10.1007/JHEP05(2023)032}{\emph{JHEP}
  {\bfseries 05} (2023) 032}
  [\href{https://arxiv.org/abs/2107.09062}{{\ttfamily 2107.09062}}].

\bibitem{Cabo-Bizet:2018ehj}
A.~Cabo-Bizet, D.~Cassani, D.~Martelli and S.~Murthy, \emph{{Microscopic origin
  of the Bekenstein-Hawking entropy of supersymmetric AdS$_{5}$ black holes}},
  \href{https://doi.org/10.1007/JHEP10(2019)062}{\emph{JHEP} {\bfseries 10}
  (2019) 062} [\href{https://arxiv.org/abs/1810.11442}{{\ttfamily
  1810.11442}}].

\bibitem{Anupam:2023yns}
A.H.~Anupam, C.~Chowdhury and A.~Sen, \emph{{Revisiting logarithmic correction
  to five dimensional BPS black hole entropy}},
  \href{https://doi.org/10.1007/JHEP05(2024)070}{\emph{JHEP} {\bfseries 05}
  (2024) 070} [\href{https://arxiv.org/abs/2308.00038}{{\ttfamily
  2308.00038}}].

\bibitem{Aharony:2021zkr}
O.~Aharony, F.~Benini, O.~Mamroud and E.~Milan, \emph{{A gravity interpretation
  for the Bethe Ansatz expansion of the $\mathcal{N}=4$ SYM index}},
  \href{https://doi.org/10.1103/PhysRevD.104.086026}{\emph{Phys. Rev. D}
  {\bfseries 104} (2021) 086026}
  [\href{https://arxiv.org/abs/2104.13932}{{\ttfamily 2104.13932}}].

\bibitem{Hosseini:2017mds}
S.M.~Hosseini, K.~Hristov and A.~Zaffaroni, \emph{{An extremization principle
  for the entropy of rotating BPS black holes in AdS$_{5}$}},
  \href{https://doi.org/10.1007/JHEP07(2017)106}{\emph{JHEP} {\bfseries 07}
  (2017) 106} [\href{https://arxiv.org/abs/1705.05383}{{\ttfamily
  1705.05383}}].

\bibitem{Hosseini:2018dob}
S.M.~Hosseini, K.~Hristov and A.~Zaffaroni, \emph{{A note on the entropy of
  rotating BPS AdS$_7\times S^4$ black holes}},
  \href{https://doi.org/10.1007/JHEP05(2018)121}{\emph{JHEP} {\bfseries 05}
  (2018) 121} [\href{https://arxiv.org/abs/1803.07568}{{\ttfamily
  1803.07568}}].

\bibitem{Choi:2018fdc}
S.~Choi, C.~Hwang, S.~Kim and J.~Nahmgoong, \emph{{Entropy Functions of BPS
  Black Holes in AdS$_{4}$ and AdS$_{6}$}},
  \href{https://doi.org/10.3938/jkps.76.101}{\emph{J. Korean Phys. Soc.}
  {\bfseries 76} (2020) 101}
  [\href{https://arxiv.org/abs/1811.02158}{{\ttfamily 1811.02158}}].

\bibitem{Cassani:2019mms}
D.~Cassani and L.~Papini, \emph{{The BPS limit of rotating AdS black hole
  thermodynamics}}, \href{https://doi.org/10.1007/JHEP09(2019)079}{\emph{JHEP}
  {\bfseries 09} (2019) 079}
  [\href{https://arxiv.org/abs/1906.10148}{{\ttfamily 1906.10148}}].

\bibitem{Kantor:2019lfo}
G.~K\'antor, C.~Papageorgakis and P.~Richmond, \emph{{AdS$_7$ Black-Hole
  Entropy and 5D $\mathcal{N}=2$ Yang-Mills}},
  \href{https://arxiv.org/abs/1907.02923}{{\ttfamily 1907.02923}}.

\bibitem{Bobev:2019zmz}
N.~Bobev and P.M.~Crichigno, \emph{{Universal spinning black holes and theories
  of class $ \mathcal{R} $}},
  \href{https://doi.org/10.1007/JHEP12(2019)054}{\emph{JHEP} {\bfseries 12}
  (2019) 054} [\href{https://arxiv.org/abs/1909.05873}{{\ttfamily
  1909.05873}}].

\bibitem{Bobev:2020pjk}
N.~Bobev, A.M.~Charles and V.S.~Min, \emph{{Euclidean black saddles and
  AdS$_{4}$ black holes}},
  \href{https://doi.org/10.1007/JHEP10(2020)073}{\emph{JHEP} {\bfseries 10}
  (2020) 073} [\href{https://arxiv.org/abs/2006.01148}{{\ttfamily
  2006.01148}}].

\bibitem{Larsen:2021wnu}
F.~Larsen and S.~Lee, \emph{{Microscopic entropy of AdS$_{3}$ black holes
  revisited}}, \href{https://doi.org/10.1007/JHEP07(2021)038}{\emph{JHEP}
  {\bfseries 07} (2021) 038}
  [\href{https://arxiv.org/abs/2101.08497}{{\ttfamily 2101.08497}}].

\bibitem{BenettiGenolini:2023ucp}
P.~Benetti~Genolini and C.~Toldo, \emph{{Magnetic charge and black hole
  supersymmetric quantum statistical relation}},
  \href{https://doi.org/10.1103/PhysRevD.107.L121902}{\emph{Phys. Rev. D}
  {\bfseries 107} (2023) L121902}
  [\href{https://arxiv.org/abs/2304.00605}{{\ttfamily 2304.00605}}].

\bibitem{Hristov:2022pmo}
K.~Hristov, \emph{{The dark (BPS) side of thermodynamics in Minkowski$_{4}$}},
  \href{https://doi.org/10.1007/JHEP09(2022)204}{\emph{JHEP} {\bfseries 09}
  (2022) 204} [\href{https://arxiv.org/abs/2207.12437}{{\ttfamily
  2207.12437}}].

\bibitem{H:2023qko}
A.A.~H., P.V.~Athira, C.~Chowdhury and A.~Sen, \emph{{Logarithmic correction to
  BPS black hole entropy from supersymmetric index at finite temperature}},
  \href{https://doi.org/10.1007/JHEP03(2024)095}{\emph{JHEP} {\bfseries 03}
  (2024) 095} [\href{https://arxiv.org/abs/2306.07322}{{\ttfamily
  2306.07322}}].

\bibitem{Boruch:2023gfn}
J.~Boruch, L.V.~Iliesiu, S.~Murthy and G.J.~Turiaci, \emph{{New forms of
  attraction: Attractor saddles for the black hole index}},
  \href{https://arxiv.org/abs/2310.07763}{{\ttfamily 2310.07763}}.

\bibitem{Cassani:2024kjn}
D.~Cassani, A.~Ruip\'erez and E.~Turetta, \emph{{Localization of the 5D
  supergravity action and Euclidean saddles for the black hole index}},
  \href{https://doi.org/10.1007/JHEP12(2024)086}{\emph{JHEP} {\bfseries 12}
  (2024) 086} [\href{https://arxiv.org/abs/2409.01332}{{\ttfamily
  2409.01332}}].

\bibitem{Boruch:2025qdq}
J.~Boruch, R.~Emparan, L.V.~Iliesiu and S.~Murthy, \emph{{The gravitational
  index of 5d black holes and black strings}},
  \href{https://arxiv.org/abs/2501.17909}{{\ttfamily 2501.17909}}.

\bibitem{Cassani:2021dwa}
D.~Cassani, J.P.~Gauntlett, D.~Martelli and J.~Sparks, \emph{{Thermodynamics of
  accelerating and supersymmetric AdS4 black holes}},
  \href{https://doi.org/10.1103/PhysRevD.104.086005}{\emph{Phys. Rev. D}
  {\bfseries 104} (2021) 086005}
  [\href{https://arxiv.org/abs/2106.05571}{{\ttfamily 2106.05571}}].

\bibitem{Benini:2015eyy}
F.~Benini, K.~Hristov and A.~Zaffaroni, \emph{{Black hole microstates in
  AdS$_{4}$ from supersymmetric localization}},
  \href{https://doi.org/10.1007/JHEP05(2016)054}{\emph{JHEP} {\bfseries 05}
  (2016) 054} [\href{https://arxiv.org/abs/1511.04085}{{\ttfamily
  1511.04085}}].

\bibitem{Choi:2018hmj}
S.~Choi, J.~Kim, S.~Kim and J.~Nahmgoong, \emph{{Large AdS black holes from
  QFT}},  \href{https://arxiv.org/abs/1810.12067}{{\ttfamily 1810.12067}}.

\bibitem{Benini:2018ywd}
F.~Benini and P.~Milan, \emph{{Black holes in 4d $\mathcal{N}=4$
  Super-Yang-Mills}},
  \href{https://doi.org/10.1103/PhysRevX.10.021037}{\emph{Phys. Rev. X}
  {\bfseries 10} (2020) 021037}
  [\href{https://arxiv.org/abs/1812.09613}{{\ttfamily 1812.09613}}].

\bibitem{Zaffaroni:2019dhb}
A.~Zaffaroni, \emph{{Lectures on AdS Black Holes, Holography and
  Localization}},  2, 2019 [\href{https://arxiv.org/abs/1902.07176}{{\ttfamily
  1902.07176}}].

\bibitem{Gunaydin:1984ak}
M.~Gunaydin, G.~Sierra and P.K.~Townsend, \emph{{Gauging the d = 5
  Maxwell-Einstein Supergravity Theories: More on Jordan Algebras}},
  \href{https://doi.org/10.1016/0550-3213(85)90547-4}{\emph{Nucl. Phys. B}
  {\bfseries 253} (1985) 573}.

\bibitem{Chong:2005hr}
Z.W.~Chong, M.~Cvetic, H.~Lu and C.N.~Pope, \emph{{General non-extremal
  rotating black holes in minimal five-dimensional gauged supergravity}},
  \href{https://doi.org/10.1103/PhysRevLett.95.161301}{\emph{Phys. Rev. Lett.}
  {\bfseries 95} (2005) 161301}
  [\href{https://arxiv.org/abs/hep-th/0506029}{{\ttfamily hep-th/0506029}}].

\bibitem{Gutowski:2004ez}
J.B.~Gutowski and H.S.~Reall, \emph{{Supersymmetric AdS$_5$ black holes}},
  \href{https://doi.org/10.1088/1126-6708/2004/02/006}{\emph{JHEP} {\bfseries
  02} (2004) 006} [\href{https://arxiv.org/abs/hep-th/0401042}{{\ttfamily
  hep-th/0401042}}].

\bibitem{Romelsberger:2005eg}
C.~Romelsberger, \emph{{Counting chiral primaries in N = 1, d=4 superconformal
  field theories}},
  \href{https://doi.org/10.1016/j.nuclphysb.2006.03.037}{\emph{Nucl. Phys.}
  {\bfseries B747} (2006) 329}
  [\href{https://arxiv.org/abs/hep-th/0510060}{{\ttfamily hep-th/0510060}}].

\bibitem{Kinney:2005ej}
J.~Kinney, J.M.~Maldacena, S.~Minwalla and S.~Raju, \emph{{An Index for 4
  dimensional super conformal theories}},
  \href{https://doi.org/10.1007/s00220-007-0258-7}{\emph{Commun. Math. Phys.}
  {\bfseries 275} (2007) 209}
  [\href{https://arxiv.org/abs/hep-th/0510251}{{\ttfamily hep-th/0510251}}].

\bibitem{GonzalezLezcano:2019nca}
A.~Gonz\'alez~Lezcano and L.A.~Pando~Zayas, \emph{{Microstate counting via
  Bethe Ans\"atze in the 4d $ \mathcal{N} $ = 1 superconformal index}},
  \href{https://doi.org/10.1007/JHEP03(2020)088}{\emph{JHEP} {\bfseries 03}
  (2020) 088} [\href{https://arxiv.org/abs/1907.12841}{{\ttfamily
  1907.12841}}].

\bibitem{Lanir:2019abx}
A.~Lanir, A.~Nedelin and O.~Sela, \emph{{Black hole entropy function for toric
  theories via Bethe Ansatz}},
  \href{https://doi.org/10.1007/JHEP04(2020)091}{\emph{JHEP} {\bfseries 04}
  (2020) 091} [\href{https://arxiv.org/abs/1908.01737}{{\ttfamily
  1908.01737}}].

\bibitem{Benini:2020gjh}
F.~Benini, E.~Colombo, S.~Soltani, A.~Zaffaroni and Z.~Zhang,
  \emph{{Superconformal indices at large $N$ and the entropy of AdS$_5$
  $\times$ SE$_5$ black holes}},
  \href{https://doi.org/10.1088/1361-6382/abb39b}{\emph{Class. Quant. Grav.}
  {\bfseries 37} (2020) 215021}
  [\href{https://arxiv.org/abs/2005.12308}{{\ttfamily 2005.12308}}].

\bibitem{Cabo-Bizet:2019eaf}
A.~Cabo-Bizet and S.~Murthy, \emph{{Supersymmetric phases of 4d $ \mathcal{N} $
  = 4 SYM at large $N$}},
  \href{https://doi.org/10.1007/JHEP09(2020)184}{\emph{JHEP} {\bfseries 09}
  (2020) 184} [\href{https://arxiv.org/abs/1909.09597}{{\ttfamily
  1909.09597}}].

\bibitem{Cabo-Bizet:2020nkr}
A.~Cabo-Bizet, D.~Cassani, D.~Martelli and S.~Murthy, \emph{{The large-$N$
  limit of the 4d $ \mathcal{N} $ = 1 superconformal index}},
  \href{https://doi.org/10.1007/JHEP11(2020)150}{\emph{JHEP} {\bfseries 11}
  (2020) 150} [\href{https://arxiv.org/abs/2005.10654}{{\ttfamily
  2005.10654}}].

\bibitem{Cabo-Bizet:2020ewf}
A.~Cabo-Bizet, \emph{{From multi-gravitons to Black holes: The role of complex
  saddles}},  \href{https://arxiv.org/abs/2012.04815}{{\ttfamily 2012.04815}}.

\bibitem{Colombo:2021kbb}
E.~Colombo, \emph{{The large-N limit of 4d superconformal indices for general
  BPS charges}}, \href{https://doi.org/10.1007/JHEP12(2022)013}{\emph{JHEP}
  {\bfseries 12} (2022) 013}
  [\href{https://arxiv.org/abs/2110.01911}{{\ttfamily 2110.01911}}].

\bibitem{Copetti:2020dil}
C.~Copetti, A.~Grassi, Z.~Komargodski and L.~Tizzano, \emph{{Delayed
  deconfinement and the Hawking-Page transition}},
  \href{https://doi.org/10.1007/JHEP04(2022)132}{\emph{JHEP} {\bfseries 04}
  (2022) 132} [\href{https://arxiv.org/abs/2008.04950}{{\ttfamily
  2008.04950}}].

\bibitem{Choi:2021rxi}
S.~Choi, S.~Jeong, S.~Kim and E.~Lee, \emph{{Exact QFT duals of AdS black
  holes}}, \href{https://doi.org/10.1007/JHEP09(2023)138}{\emph{JHEP}
  {\bfseries 09} (2023) 138}
  [\href{https://arxiv.org/abs/2111.10720}{{\ttfamily 2111.10720}}].

\bibitem{Choi:2023tiq}
S.~Choi, S.~Kim and J.~Song, \emph{{Large N universality of 4d $ \mathcal{N} $
  = 1 superconformal index and AdS black holes}},
  \href{https://doi.org/10.1007/JHEP08(2024)105}{\emph{JHEP} {\bfseries 08}
  (2024) 105} [\href{https://arxiv.org/abs/2309.07614}{{\ttfamily
  2309.07614}}].

\bibitem{DiPietro:2014bca}
L.~Di~Pietro and Z.~Komargodski, \emph{{Cardy formulae for SUSY theories in $d
  =$ 4 and $d =$ 6}},
  \href{https://doi.org/10.1007/JHEP12(2014)031}{\emph{JHEP} {\bfseries 12}
  (2014) 031} [\href{https://arxiv.org/abs/1407.6061}{{\ttfamily 1407.6061}}].

\bibitem{Honda:2019cio}
M.~Honda, \emph{{Quantum Black Hole Entropy from 4d Supersymmetric Cardy
  formula}}, \href{https://doi.org/10.1103/PhysRevD.100.026008}{\emph{Phys.
  Rev. D} {\bfseries 100} (2019) 026008}
  [\href{https://arxiv.org/abs/1901.08091}{{\ttfamily 1901.08091}}].

\bibitem{ArabiArdehali:2019tdm}
A.~Arabi~Ardehali, \emph{{Cardy-like asymptotics of the 4d $ \mathcal{N}=4 $
  index and AdS$_{5}$ blackholes}},
  \href{https://doi.org/10.1007/JHEP06(2019)134}{\emph{JHEP} {\bfseries 06}
  (2019) 134} [\href{https://arxiv.org/abs/1902.06619}{{\ttfamily
  1902.06619}}].

\bibitem{Kim:2019yrz}
J.~Kim, S.~Kim and J.~Song, \emph{{A 4d $ \mathcal{N} $ = 1 Cardy Formula}},
  \href{https://doi.org/10.1007/JHEP01(2021)025}{\emph{JHEP} {\bfseries 01}
  (2021) 025} [\href{https://arxiv.org/abs/1904.03455}{{\ttfamily
  1904.03455}}].

\bibitem{Amariti:2019mgp}
A.~Amariti, I.~Garozzo and G.~Lo~Monaco, \emph{{Entropy function from toric
  geometry}},
  \href{https://doi.org/10.1016/j.nuclphysb.2021.115571}{\emph{Nucl. Phys. B}
  {\bfseries 973} (2021) 115571}
  [\href{https://arxiv.org/abs/1904.10009}{{\ttfamily 1904.10009}}].

\bibitem{Cabo-Bizet:2019osg}
A.~Cabo-Bizet, D.~Cassani, D.~Martelli and S.~Murthy, \emph{{The asymptotic
  growth of states of the 4d $ \mathcal{N}=1 $ superconformal index}},
  \href{https://doi.org/10.1007/JHEP08(2019)120}{\emph{JHEP} {\bfseries 08}
  (2019) 120} [\href{https://arxiv.org/abs/1904.05865}{{\ttfamily
  1904.05865}}].

\bibitem{Gadde:2020bov}
A.~Gadde, \emph{{Modularity of supersymmetric partition functions}},
  \href{https://arxiv.org/abs/2004.13490}{{\ttfamily 2004.13490}}.

\bibitem{GonzalezLezcano:2020yeb}
A.~Gonz\'alez~Lezcano, J.~Hong, J.T.~Liu and L.A.~Pando~Zayas,
  \emph{{Sub-leading Structures in Superconformal Indices: Subdominant Saddles
  and Logarithmic Contributions}},
  \href{https://doi.org/10.1007/JHEP01(2021)001}{\emph{JHEP} {\bfseries 01}
  (2021) 001} [\href{https://arxiv.org/abs/2007.12604}{{\ttfamily
  2007.12604}}].

\bibitem{Goldstein:2020yvj}
K.~Goldstein, V.~Jejjala, Y.~Lei, S.~van Leuven and W.~Li, \emph{{Residues,
  modularity, and the Cardy limit of the 4d $ \mathcal{N} $ = 4 superconformal
  index}}, \href{https://doi.org/10.1007/JHEP04(2021)216}{\emph{JHEP}
  {\bfseries 04} (2021) 216}
  [\href{https://arxiv.org/abs/2011.06605}{{\ttfamily 2011.06605}}].

\bibitem{Amariti:2020jyx}
A.~Amariti, M.~Fazzi and A.~Segati, \emph{{The SCI of $ \mathcal{N} $ = 4
  USp(2N$_{c}$) and SO(N$_{c}$) SYM as a matrix integral}},
  \href{https://doi.org/10.1007/JHEP06(2021)132}{\emph{JHEP} {\bfseries 06}
  (2021) 132} [\href{https://arxiv.org/abs/2012.15208}{{\ttfamily
  2012.15208}}].

\bibitem{Amariti:2021ubd}
A.~Amariti, M.~Fazzi and A.~Segati, \emph{{Expanding on the Cardy-like limit of
  the SCI of 4d $ \mathcal{N} $ = 1 ABCD SCFTs}},
  \href{https://doi.org/10.1007/JHEP07(2021)141}{\emph{JHEP} {\bfseries 07}
  (2021) 141} [\href{https://arxiv.org/abs/2103.15853}{{\ttfamily
  2103.15853}}].

\bibitem{Cassani:2021fyv}
D.~Cassani and Z.~Komargodski, \emph{{EFT and the SUSY Index on the 2nd
  Sheet}}, \href{https://doi.org/10.21468/SciPostPhys.11.1.004}{\emph{SciPost
  Phys.} {\bfseries 11} (2021) 004}
  [\href{https://arxiv.org/abs/2104.01464}{{\ttfamily 2104.01464}}].

\bibitem{ArabiArdehali:2021nsx}
A.~Arabi~Ardehali and S.~Murthy, \emph{{The 4d superconformal index near roots
  of unity and 3d Chern-Simons theory}},
  \href{https://doi.org/10.1007/JHEP10(2021)207}{\emph{JHEP} {\bfseries 10}
  (2021) 207} [\href{https://arxiv.org/abs/2104.02051}{{\ttfamily
  2104.02051}}].

\bibitem{Cassani:2022lrk}
D.~Cassani, A.~Ruip\'erez and E.~Turetta, \emph{{Corrections to AdS$_{5}$ black
  hole thermodynamics from higher-derivative supergravity}},
  \href{https://doi.org/10.1007/JHEP11(2022)059}{\emph{JHEP} {\bfseries 11}
  (2022) 059} [\href{https://arxiv.org/abs/2208.01007}{{\ttfamily
  2208.01007}}].

\bibitem{Bobev:2021qxx}
N.~Bobev, K.~Hristov and V.~Reys, \emph{{AdS$_{5}$ holography and
  higher-derivative supergravity}},
  \href{https://doi.org/10.1007/JHEP04(2022)088}{\emph{JHEP} {\bfseries 04}
  (2022) 088} [\href{https://arxiv.org/abs/2112.06961}{{\ttfamily
  2112.06961}}].

\bibitem{Bobev:2022bjm}
N.~Bobev, V.~Dimitrov, V.~Reys and A.~Vekemans, \emph{{Higher derivative
  corrections and AdS5 black holes}},
  \href{https://doi.org/10.1103/PhysRevD.106.L121903}{\emph{Phys. Rev. D}
  {\bfseries 106} (2022) L121903}
  [\href{https://arxiv.org/abs/2207.10671}{{\ttfamily 2207.10671}}].

\bibitem{Witten:1998qj}
E.~Witten, \emph{{Anti-de Sitter space and holography}},
  \href{https://doi.org/10.4310/ATMP.1998.v2.n2.a2}{\emph{Adv. Theor. Math.
  Phys.} {\bfseries 2} (1998) 253}
  [\href{https://arxiv.org/abs/hep-th/9802150}{{\ttfamily hep-th/9802150}}].

\bibitem{Hanaki:2006pj}
K.~Hanaki, K.~Ohashi and Y.~Tachikawa, \emph{{Supersymmetric Completion of an
  R**2 term in Five-dimensional Supergravity}},
  \href{https://doi.org/10.1143/PTP.117.533}{\emph{Prog. Theor. Phys.}
  {\bfseries 117} (2007) 533}
  [\href{https://arxiv.org/abs/hep-th/0611329}{{\ttfamily hep-th/0611329}}].

\bibitem{Ozkan:2013nwa}
M.~Ozkan and Y.~Pang, \emph{{All off-shell $R^{2}$ invariants in five
  dimensional $\mathcal{N} =$ 2 supergravity}},
  \href{https://doi.org/10.1007/JHEP08(2013)042}{\emph{JHEP} {\bfseries 08}
  (2013) 042} [\href{https://arxiv.org/abs/1306.1540}{{\ttfamily 1306.1540}}].

\bibitem{Gold:2023ymc}
G.~Gold, J.~Hutomo, S.~Khandelwal, M.~Ozkan, Y.~Pang and
  G.~Tartaglino-Mazzucchelli, \emph{{All Gauged Curvature-Squared
  Supergravities in Five Dimensions}},
  \href{https://doi.org/10.1103/PhysRevLett.131.251603}{\emph{Phys. Rev. Lett.}
  {\bfseries 131} (2023) 251603}
  [\href{https://arxiv.org/abs/2309.07637}{{\ttfamily 2309.07637}}].

\bibitem{Cassani:2023vsa}
D.~Cassani, A.~Ruip\'erez and E.~Turetta, \emph{{Boundary terms and conserved
  charges in higher-derivative gauged supergravity}},
  \href{https://doi.org/10.1007/JHEP06(2023)203}{\emph{JHEP} {\bfseries 06}
  (2023) 203} [\href{https://arxiv.org/abs/2304.06101}{{\ttfamily
  2304.06101}}].

\bibitem{Cano:2024tcr}
P.A.~Cano and M.~David, \emph{{Near-horizon geometries and black hole
  thermodynamics in higher-derivative AdS$_{5}$ supergravity}},
  \href{https://doi.org/10.1007/JHEP03(2024)036}{\emph{JHEP} {\bfseries 03}
  (2024) 036} [\href{https://arxiv.org/abs/2402.02215}{{\ttfamily
  2402.02215}}].

\bibitem{Cassani:2024tvk}
D.~Cassani, A.~Ruip\'erez and E.~Turetta, \emph{{Higher-derivative corrections
  to flavoured BPS black hole thermodynamics and holography}},
  \href{https://doi.org/10.1007/JHEP05(2024)276}{\emph{JHEP} {\bfseries 05}
  (2024) 276} [\href{https://arxiv.org/abs/2403.02410}{{\ttfamily
  2403.02410}}].

\bibitem{Bobev:2021oku}
N.~Bobev, A.M.~Charles, K.~Hristov and V.~Reys, \emph{{Higher-derivative
  supergravity, AdS$_{4}$ holography, and black holes}},
  \href{https://doi.org/10.1007/JHEP08(2021)173}{\emph{JHEP} {\bfseries 08}
  (2021) 173} [\href{https://arxiv.org/abs/2106.04581}{{\ttfamily
  2106.04581}}].

\bibitem{BenettiGenolini:2023rkq}
P.~Benetti~Genolini, A.~Cabo-Bizet and S.~Murthy, \emph{{Supersymmetric phases
  of AdS$_{4}$/CFT$_{3}$}},
  \href{https://doi.org/10.1007/JHEP06(2023)125}{\emph{JHEP} {\bfseries 06}
  (2023) 125} [\href{https://arxiv.org/abs/2301.00763}{{\ttfamily
  2301.00763}}].

\bibitem{Aharony:2024ntg}
O.~Aharony, O.~Mamroud, S.~Nowik and M.~Weissman, \emph{{Bethe Ansatz for the
  superconformal index with unequal angular momenta}},
  \href{https://doi.org/10.1103/PhysRevD.109.085015}{\emph{Phys. Rev. D}
  {\bfseries 109} (2024) 085015}
  [\href{https://arxiv.org/abs/2402.03977}{{\ttfamily 2402.03977}}].

\bibitem{Cvetic:1996xz}
M.~Cvetic and D.~Youm, \emph{{General rotating five-dimensional black holes of
  toroidally compactified heterotic string}},
  \href{https://doi.org/10.1016/0550-3213(96)00355-0}{\emph{Nucl. Phys. B}
  {\bfseries 476} (1996) 118}
  [\href{https://arxiv.org/abs/hep-th/9603100}{{\ttfamily hep-th/9603100}}].

\bibitem{Breckenridge:1996is}
J.C.~Breckenridge, R.C.~Myers, A.W.~Peet and C.~Vafa, \emph{{D-branes and
  spinning black holes}},
  \href{https://doi.org/10.1016/S0370-2693(96)01460-8}{\emph{Phys. Lett. B}
  {\bfseries 391} (1997) 93}
  [\href{https://arxiv.org/abs/hep-th/9602065}{{\ttfamily hep-th/9602065}}].

\bibitem{Hegde:2023jmp}
S.~Hegde and A.~Virmani, \emph{{Killing spinors for finite temperature
  Euclidean solutions at the BPS bound}},
  \href{https://doi.org/10.1007/JHEP02(2024)203}{\emph{JHEP} {\bfseries 02}
  (2024) 203} [\href{https://arxiv.org/abs/2311.09427}{{\ttfamily
  2311.09427}}].

\bibitem{Israel:1972vx}
W.~Israel and G.A.~Wilson, \emph{{A class of stationary electromagnetic vacuum
  fields}}, \href{https://doi.org/10.1063/1.1666066}{\emph{J. Math. Phys.}
  {\bfseries 13} (1972) 865}.

\bibitem{Perjes:1971gv}
Z.~Perjes, \emph{{Solutions of the coupled Einstein Maxwell equations
  representing the fields of spinning sources}},
  \href{https://doi.org/10.1103/PhysRevLett.27.1668}{\emph{Phys. Rev. Lett.}
  {\bfseries 27} (1971) 1668}.

\bibitem{Whitt:1984wk}
B.~Whitt, \emph{{Israel-Wilson Metrics}},
  \href{https://doi.org/10.1016/0003-4916(85)90079-X}{\emph{Annals Phys.}
  {\bfseries 161} (1985) 241}.

\bibitem{Yuille:1987vw}
A.L.~Yuille, \emph{{Israel-wilson Metrics in the Euclidean Regime}},
  \href{https://doi.org/10.1088/0264-9381/4/5/034}{\emph{Class. Quant. Grav.}
  {\bfseries 4} (1987) 1409}.

\bibitem{BenettiGenolini:2023kxp}
P.~Benetti~Genolini, J.P.~Gauntlett and J.~Sparks, \emph{{Equivariant
  Localization in Supergravity}},
  \href{https://doi.org/10.1103/PhysRevLett.131.121602}{\emph{Phys. Rev. Lett.}
  {\bfseries 131} (2023) 121602}
  [\href{https://arxiv.org/abs/2306.03868}{{\ttfamily 2306.03868}}].

\bibitem{Martelli:2023oqk}
D.~Martelli and A.~Zaffaroni, \emph{{Equivariant localization and holography}},
  \href{https://doi.org/10.1007/s11005-023-01752-1}{\emph{Lett. Math. Phys.}
  {\bfseries 114} (2024) 15}
  [\href{https://arxiv.org/abs/2306.03891}{{\ttfamily 2306.03891}}].

\bibitem{BenettiGenolini:2023ndb}
P.~Benetti~Genolini, J.P.~Gauntlett and J.~Sparks, \emph{{Equivariant
  localization for AdS/CFT}},
  \href{https://doi.org/10.1007/JHEP02(2024)015}{\emph{JHEP} {\bfseries 02}
  (2024) 015} [\href{https://arxiv.org/abs/2308.11701}{{\ttfamily
  2308.11701}}].

\bibitem{BenettiGenolini:2024kyy}
P.~Benetti~Genolini, J.P.~Gauntlett, Y.~Jiao, A.~L\"uscher and J.~Sparks,
  \emph{{Localization and attraction}},
  \href{https://doi.org/10.1007/JHEP05(2024)152}{\emph{JHEP} {\bfseries 05}
  (2024) 152} [\href{https://arxiv.org/abs/2401.10977}{{\ttfamily
  2401.10977}}].

\bibitem{Gibbons:1979xm}
G.W.~Gibbons and S.W.~Hawking, \emph{{Classification of Gravitational Instanton
  Symmetries}}, \href{https://doi.org/10.1007/BF01197189}{\emph{Commun. Math.
  Phys.} {\bfseries 66} (1979) 291}.

\bibitem{Chen:2023lzq}
Y.~Chen, M.~Heydeman, Y.~Wang and M.~Zhang, \emph{{Probing supersymmetric black
  holes with surface defects}},
  \href{https://doi.org/10.1007/JHEP10(2023)136}{\emph{JHEP} {\bfseries 10}
  (2023) 136} [\href{https://arxiv.org/abs/2306.05463}{{\ttfamily
  2306.05463}}].

\bibitem{Kontsevich:2021dmb}
M.~Kontsevich and G.~Segal, \emph{{Wick Rotation and the Positivity of Energy
  in Quantum Field Theory}},
  \href{https://doi.org/10.1093/qmath/haab027}{\emph{Quart. J. Math. Oxford
  Ser.} {\bfseries 72} (2021) 673}
  [\href{https://arxiv.org/abs/2105.10161}{{\ttfamily 2105.10161}}].

\bibitem{Witten:2021nzp}
E.~Witten, \emph{{A Note On Complex Spacetime Metrics}},
  \href{https://arxiv.org/abs/2111.06514}{{\ttfamily 2111.06514}}.

\bibitem{Sen:2012dw}
A.~Sen, \emph{{Logarithmic Corrections to Schwarzschild and Other Non-extremal
  Black Hole Entropy in Different Dimensions}},
  \href{https://doi.org/10.1007/JHEP04(2013)156}{\emph{JHEP} {\bfseries 04}
  (2013) 156} [\href{https://arxiv.org/abs/1205.0971}{{\ttfamily 1205.0971}}].

\bibitem{Gupta:2021roy}
R.K.~Gupta, S.~Murthy and M.~Sahni, \emph{{Quantum entropy of BMPV black holes
  and the topological M-theory conjecture}},
  \href{https://doi.org/10.1007/JHEP06(2022)053}{\emph{JHEP} {\bfseries 06}
  (2022) 053} [\href{https://arxiv.org/abs/2104.02634}{{\ttfamily
  2104.02634}}].

\bibitem{Dabholkar:2012nd}
A.~Dabholkar, S.~Murthy and D.~Zagier, \emph{{Quantum Black Holes, Wall
  Crossing, and Mock Modular Forms}},
  \href{https://arxiv.org/abs/1208.4074}{{\ttfamily 1208.4074}}.

\bibitem{Ferrari:2017msn}
F.~Ferrari and V.~Reys, \emph{{Mixed Rademacher and BPS Black Holes}},
  \href{https://doi.org/10.1007/JHEP07(2017)094}{\emph{JHEP} {\bfseries 07}
  (2017) 094} [\href{https://arxiv.org/abs/1702.02755}{{\ttfamily
  1702.02755}}].

\bibitem{Bhand:2023rhm}
A.~Bhand, A.~Sen and R.K.~Singh, \emph{{Mock modularity in CHL models}},
  \href{https://doi.org/10.1007/s40687-024-00489-0}{\emph{Res. Math. Sci.}
  {\bfseries 12} (2025) 8} [\href{https://arxiv.org/abs/2311.16252}{{\ttfamily
  2311.16252}}].

\bibitem{Denef:2007vg}
F.~Denef and G.W.~Moore, \emph{{Split states, entropy enigmas, holes and
  halos}}, \href{https://doi.org/10.1007/JHEP11(2011)129}{\emph{JHEP}
  {\bfseries 11} (2011) 129}
  [\href{https://arxiv.org/abs/hep-th/0702146}{{\ttfamily hep-th/0702146}}].

\bibitem{Dabholkar:2014wpa}
A.~Dabholkar, N.~Drukker and J.~Gomes, \emph{{Localization in supergravity and
  quantum $AdS_4/CFT_3$ holography}},
  \href{https://doi.org/10.1007/JHEP10(2014)090}{\emph{JHEP} {\bfseries 10}
  (2014) 090} [\href{https://arxiv.org/abs/1406.0505}{{\ttfamily 1406.0505}}].

\bibitem{Hristov:2018lod}
K.~Hristov, I.~Lodato and V.~Reys, \emph{{On the quantum entropy function in 4d
  gauged supergravity}},
  \href{https://doi.org/10.1007/JHEP07(2018)072}{\emph{JHEP} {\bfseries 07}
  (2018) 072} [\href{https://arxiv.org/abs/1803.05920}{{\ttfamily
  1803.05920}}].

\bibitem{Hristov:2021zai}
K.~Hristov and V.~Reys, \emph{{Factorization of log-corrections in
  AdS$_{4}$/CFT$_{3}$ from supergravity localization}},
  \href{https://doi.org/10.1007/JHEP12(2021)031}{\emph{JHEP} {\bfseries 12}
  (2021) 031} [\href{https://arxiv.org/abs/2107.12398}{{\ttfamily
  2107.12398}}].

\end{thebibliography}\endgroup
\bibliographystyle{JHEP}

\end{document}